\documentclass{jfm}

\usepackage{graphicx}
\usepackage{newtxtext}
\usepackage{newtxmath}
\usepackage{natbib}
\usepackage{hyperref}
\hypersetup{
    colorlinks = true,
    urlcolor   = blue,
    citecolor  = black,
}
\usepackage{caption}
\newcommand{\RomanNumeralCaps}[1]
\linenumbers
\usepackage{algorithm}
\usepackage{algpseudocode}

\title{Multiscale Hypersonic Boundary Layer Reconstruction via Spectral Binning and Subdomain-wise Conditional Diffusion}

\author{Hojin Kim\aff{1},
  Dibyajyoti Chakraborty\aff{2}, 
  Takahiko Toki\aff{1},
  Carlo Scalo\aff{1},
 \and Romit Maulik\aff{1, 3}}

\affiliation{\aff{1}School of Mechanical Engineering, Purdue University, West Lafayette, 47907, IN, USA
\aff{2}College of Information Sciences and Technology, Pennsylvania State University, University Park, 16802, PA, USA
\aff{3}Mathematics and Computer Science Division, Argonne National Laboratory, Lemont, 60439, IL, USA}

\begin{document}
\maketitle

\begin{abstract}
We propose a multiscale probabilistic reconstruction framework for hypersonic Couette flow, where near-wall states are inferred from limited top-wall observations using conditional diffusion model. The boundary layer is divided into overlapping wall-normal subdomains, and a single height- and Mach-conditioned Elucidating Diffusion Model (EDM) is trained jointly for $M=6,7,8$ to sample velocity, density, pressure, and temperature fields conditioned on a top-wall boundary slice. A soft overlap inpainting strategy assembles subdomain predictions into full-volume reconstructions while maintaining inter-subdomain continuity and small-scale variability. To improve the spectral fidelity of the generated fields, we introduce a novel bounded binned spectral power (BSP) loss that preserves high-wavenumber content while remaining numerically stable across the diffusion noise schedule. Validation against direct numerical simulation data shows that the model recovers instantaneous structures, spectra, statistical profiles, correlations, and wall quantities across all training Mach numbers, while providing spatially structured uncertainty estimates. The reconstructed Mach-conditioned profiles also collapse under the Trettel–Larsson transformation, indicating consistency with compressibility scaling. These results establish the domain decomposed conditional diffusion model with a bounded binned spectral loss as an effective probabilistic surrogate for near-wall reconstruction in hypersonic wall-bounded turbulence.
\end{abstract}

\begin{keywords}
Authors should not enter keywords on the manuscript, as these must be chosen by the author during the online submission process and will then be added during the typesetting process (see \href{https://www.cambridge.org/core/journals/journal-of-fluid-mechanics/information/list-of-keywords}{Keyword PDF} for the full list).  Other classifications will be added at the same time.
\end{keywords}

{\bf MSC Codes }  {\it(Optional)} Please enter your MSC Codes here

\section{Introduction}
\label{sec:Introduction}
High-fidelity simulation of high-speed boundary layers is notoriously computationally expensive. In wall-bounded flows, dynamically relevant structures become progressively thinner as the wall is approached, which makes accurate resolution of near-wall behavior increasingly demanding \citep{sillero2014two,bose2018wall,choi2012grid}. In hypersonic regimes, this difficulty is further amplified by compressibility effects, sharp velocity and temperature gradients, and the stringent wall-normal resolution required to capture coupled momentum and heat-transfer processes. These challenges motivate data-driven approaches that can reconstruct unresolved boundary-layer structure from limited observations while preserving the essential multiscale physics of the flow.

Boundary-layer reconstruction from sparse observations, however, is fundamentally an ill-posed inverse problem. A given set of boundary measurements may be consistent with multiple plausible near-wall states, especially in turbulent flows where fine-scale fluctuations are only partially observed. Conventional deterministic approaches can provide useful point estimates \citep{guastoni2021convolutional, guastoni2025fully, fukami2019super, fukami2021global}, but they are not naturally suited to representing the one-to-many character of this mapping. As a result, they may suppress physically relevant variability, smooth small-scale content, and provide limited information about predictive uncertainty. For high-speed boundary-layer reconstruction, where both multiscale structure and confidence in the recovered state are important, a probabilistic formulation is therefore especially attractive.

Generative models provide a natural framework for such inverse problems because they aim to learn conditional distributions rather than single deterministic outputs. Among them, diffusion models \citep{song2020score, ho2020denoising} have recently emerged as particularly promising tools for turbulent flow modeling and reconstruction. In related turbulence applications, diffusion-based approaches have been shown to improve the spectral representation of generated fields \citep{oommen2025integrating}. Conditional diffusion formulations have also been used for reconstructing high-fidelity turbulence under physics-informed conditioning \citep{shu2023physics}, latent generative modeling for turbulence inflow generation \citep{du2024confild}, and probabilistic reconstruction of turbulent flow fields \citep{amoros2026guiding}. In parallel, autoregressive conditional diffusion has been shown to provide stable probabilistic rollouts for turbulent flow simulation~\citep{kohl2026benchmarking, jiang2025integrating, chakraborty2026adaptive}, purely generative formulations have produced three-dimensional turbulence from geometry information~\citep{lienen2023zero}, and latent conditional neural-field diffusion has enabled efficient spatiotemporal turbulence generation in complex 3D domains~\citep{du2024confild}. These advances suggest that diffusion models are well suited to recovering complex, multiscale flow structure from incomplete information while also enabling uncertainty quantification through ensemble sampling.

Recent studies have further extended diffusion models specifically to wall-bounded flows. Examples include spatiotemporal turbulence generation in backward-facing step and channel flows, together with two-dimensional slices of hypersonic channel flow~\citep{gao2024bayesian}; turbulence generation and data assimilation in incompressible Couette flow~\citep{steinbrenner2026turbulence}; wall-pressure reconstruction from sparse wall sensors~\citep{fan2025generative}; and near-wall reconstruction in incompressible flows from wall measurements~\citep{parikh2025conditional}. Despite this growing body of work, existing wall-bounded studies largely focus on temporal autoregressive rollouts or on single-shot generation of 2D slices and fixed-size 3D boxes. Probabilistic reconstruction of full three-dimensional hypersonic Couette boundary-layer fields from top-wall boundary information, with explicit control over wall-normal position and compressibility through the Mach number, to our knowledge, has not been addressed.

% \noindent\textbf{Chunked and patch-wise diffusion beyond turbulence.}
A separate line of work, developed primarily in the computer-vision and medical-imaging communities, addresses a closely related question: how to generate large, coherent outputs from diffusion computations performed on smaller patches or chunks. Two families of methods have emerged. Training-free joint-diffusion approaches use a pre-trained diffusion model and reconcile overlapping views at every reverse-diffusion step: MultiDiffusion \citep{bar2023multidiffusion} stitches panoramas by arithmetic averaging of overlapping latent predictions, Mixture of Diffusers \citep{jimenez2023mixture} generalizes this with Gaussian-weighted averaging, and SyncDiffusion \citep{lee2023syncdiffusion} replaces averaging with gradient descent on a perceptual similarity loss to improve global coherence. Patch-trained diffusion approaches, in contrast, train the model directly on patches with coordinate-based conditioning: Patch Diffusion \citep{wang2023patch} introduces patch location as additional coordinate channels for efficient training on 2D images, and subsequent volumetric extensions such as PatchDDM \citep{bieder2024memory}, MedDiff-FM \citep{yu2024meddiff}, and 3D MedDiffusion \citep{wang20253d} apply patch training with coordinate conditioning and overlapping sliding-window sampling to high-resolution 3D medical volumes. Complementing these, RePaint \citep{lugmayr2022repaint} establishes a powerful inpainting-based conditioning mechanism by injecting known information into the reverse diffusion process through replacement and resampling.
 
% \noindent\textbf{Limitations of existing chunked strategies for wall-bounded turbulence.}
These approaches, while directly relevant, are not designed for the problem considered here and exhibit three limitations when transferred to hypersonic boundary-layer reconstruction. First, they partition along translationally homogeneous axes such as panoramic width or medical-volume grid coordinates, whereas subdomains along the wall-normal direction of a boundary layer span physically heterogeneous regimes (viscous sublayer, buffer, log-law, and outer regions) whose statistics change continuously with wall distance and must be learned as a function of a physically meaningful coordinate. Second, the dominant reconciliation strategies such as arithmetic or Gaussian averaging of overlapping predictions are acceptable for natural images but systematically damp the small-scale variance of turbulent fluctuations, because averaging $N$ independent stochastic samples reduces their variance by $1/N$, while the fluctuation statistics are precisely the quantity of interest for turbulence. Third, none of these frameworks addresses compressibility or multivariate thermophysical coupling, and none explicitly supervises the multiscale energy content of the generated fields, which is essential whenever the generated samples will be assessed against a Kolmogorov-like energy cascade rather than perceptual image quality.

Motivated by these gaps, we develop a multiscale probabilistic reconstruction framework for hypersonic Couette boundary layers that is organized around two complementary ideas reflected in the title: a \emph{subdomain-wise conditional diffusion} generation strategy, which decomposes the full volume into wall-normal subdomains generated by a single representative height- and Mach-conditioned diffusion model and stitched together with a continuity- and variance-preserving inpainting scheme; and a \emph{spectral-binning} training loss, which explicitly supervises binned one-dimensional energy spectra in a form that remains numerically stable across the entire diffusion noise schedule. The framework is positioned at the intersection of the lines reviewed above: it inherits the coordinate-conditioning idea from Patch Diffusion \citep{wang2023patch} and volumetric patch-trained diffusion, but replaces homogeneous-axis coordinates with physically heterogeneous wall-normal conditioning and a physical parameter (Mach number). It replaces averaging-based composition of overlapping subdomains with a replacement-based inpainting mechanism inspired by RePaint \citep{lugmayr2022repaint}, adapted with a noise-level-dependent blending schedule designed to preserve turbulent fluctuation variance rather than collapse it, and it introduces a multiscale spectral supervision specific to turbulent flows. Concretely, the main contributions are:
\begin{enumerate}
    \item \textbf{Subdomain-wise conditional diffusion for multiscale boundary layers.} A height- and Mach-conditioned EDM-based diffusion model that samples multivariate wall-normal subdomains (velocity, density, pressure, temperature) of hypersonic Couette boundary layers from a single top-wall boundary slice, making the full-volume generation tractable without collapsing the multiscale structure of the flow.
    \item \textbf{Soft overlap inpainting for subdomain composition.} A replacement-based, noise-level-dependent blending scheme that stitches subdomain-wise samples into a full volume, preserving inter-subdomain continuity while retaining the small-scale variance that is otherwise damped by the arithmetic or Gaussian averaging used in prior joint-diffusion frameworks.
    \item \textbf{Bounded spectral-binning loss.} A bounded, symmetric variant of the binned spectral power (BSP) loss that remains numerically stable across the EDM noise schedule and sharpens high-wavenumber spectral fidelity of the generated fields, which is especially important for multiscale turbulence.
    \item \textbf{Comprehensive probabilistic evaluation.} A detailed assessment on DNS of hypersonic turbulent Couette flow at $M=6,7,8$ by a single jointly-trained Mach-conditioned model, including instantaneous contours, PDFs, one-dimensional energy and Reynolds-stress spectra, mean and Reynolds-stress profiles with uncertainty bands, spatial correlations, uncertainty maps, wall shear stress and heat flux statistics, and Trettel-Larsson transformed profiles that indicate the learned Mach conditioning is consistent with known compressibility scaling of wall-bounded turbulence.
\end{enumerate}

The remainder of this paper is organized as follows. Section~\ref{sec:dataset} describes the hypersonic Couette DNS dataset used for training and evaluation. Section~\ref{sec:methods} introduces the subdomain-wise problem formulation, the EDM-based conditional diffusion model, the bounded BSP spectral loss, and the soft overlap inpainting strategy for full volume reconstruction. Section~\ref{sec:Results} presents the evaluation results, including near-wall turbulent statistics, uncertainty quantification, wall-quantity predictions, and multi-Mach performance. Section~\ref{sec:conclusions} concludes and outlines future work.

\section{Dataset}
\label{sec:dataset}

We employ the hypersonic turbulent Couette flow dataset of~\citet{toki2024sub}, which provides direct numerical simulation (DNS) data for wall-bounded hypersonic turbulence at Mach numbers $M = 6, 7, 8$. All three Mach-number cases are used jointly during training, with the Mach number supplied to the model as a conditioning variable (see Sec.~\ref{subsec:diffusion}); a single unified model is therefore obtained, and its performance is reported for each of the three training Mach numbers in Sec.~\ref{sec:Results}. We describe the $M=6$ configuration in detail below, as it is the representative case used in Sec.~\ref{subsec:training_case}. The top wall moves at $1805\,\text{m/s}$ while the bottom wall is stationary, with wall temperatures of $224\,\text{K}$ and $300\,\text{K}$, respectively. The bulk pressure is $2950\,\text{Pa}$, giving a bulk density of $0.018\,\text{kg/m}^3$. The computational domain spans $24\delta \times 2\delta \times 4\delta$ in the streamwise, wall-normal, and spanwise directions, with $\delta = 0.01\,\text{m}$. The flow is solved using the sixth-order compact finite-difference solver CFDSU~\citep{nagarajan2003robust, chen2021effects, chen2021trapped} and achieves a friction Reynolds number $Re_\tau = 621$. The DNS uses $512 \times 192 \times 256$ grid points, yielding a maximum grid spacing of $(\Delta x/\eta, \Delta y/\eta, \Delta z/\eta)_\text{max} = (20.3,\,2.1,\,6.8)$, with $\eta$ the Kolmogorov length scale. Viscous and thermal transport properties are evaluated through Sutherland's law, and both walls are maintained under isothermal no-slip conditions. The $M=7$ and $M=8$ cases share the same numerical framework and domain configuration. 

\begin{figure}
\centering
\includegraphics[width=.8\textwidth]{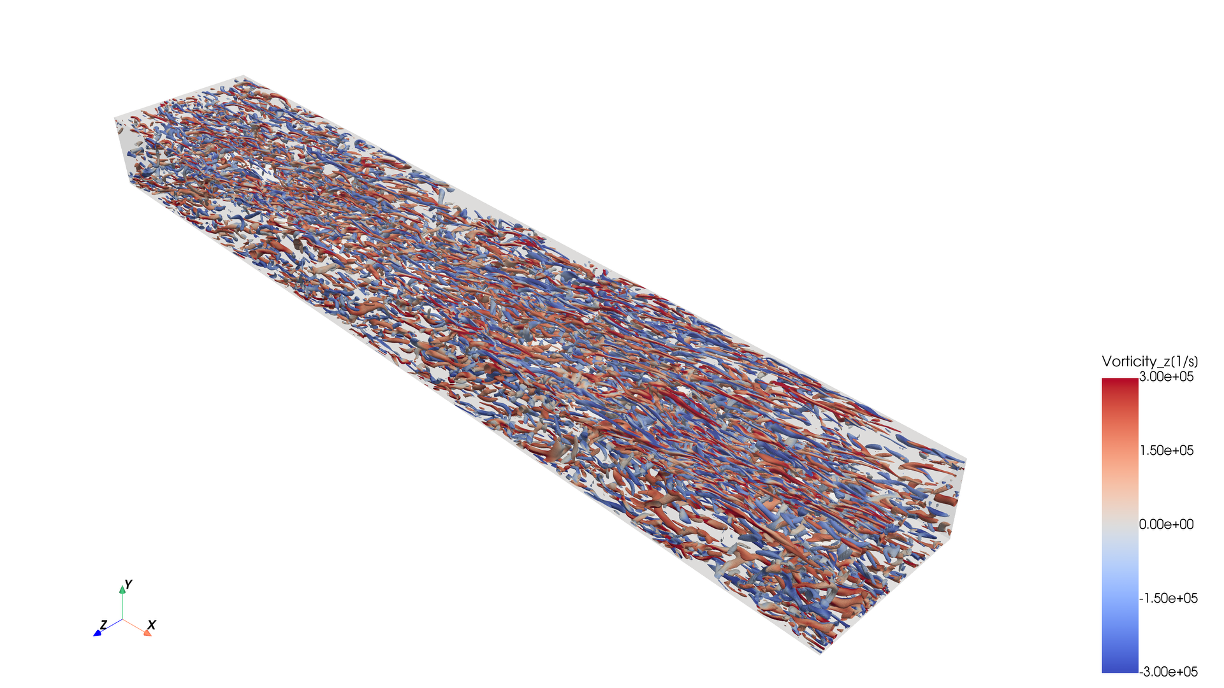}
\caption{Iso-surface of the Q-criterion colored by the spanwise vorticity component for the hypersonic turbulent Couette flow ($M=6$).}
\label{fig:couette_iso}
\end{figure}

\section{Methods}
\label{sec:methods}

% We target a probabilistic reconstruction of the full boundary-layer state from outer-layer boundary information. Section~\ref{subsec:problem} formulates the problem in a chunkwise fashion to keep the generative task tractable. Section~\ref{subsec:diffusion} details the conditional EDM framework and denoiser architecture. Section~\ref{subsec:bsp} introduces a bounded binned spectral-power (BSP) loss compatible with diffusion training. Section~\ref{subsec:fullvolume} describes how chunkwise samples are composed into a full boundary-layer volume via soft overlap inpainting.

\subsection{Problem Formulation: Subdomain-wise Boundary-Layer Reconstruction}
\label{subsec:problem}
 
Our goal is to reconstruct the hypersonic Couette boundary layer state given the flow field at the top-wall boundary and the Mach number. The dataset has dimensions $(N_x, N_y, N_z) = (512, 192, 256)$ and six flow variables $(U, V, W, \rho, P, T)$, so reconstructing the full volume in one shot would require generating a tensor in $\mathbb{R}^{6\times 512 \times 192 \times 256}$, which is computationally prohibitive for diffusion-based training and sampling.
 
\begin{figure}
\centering
\includegraphics[width=\textwidth]{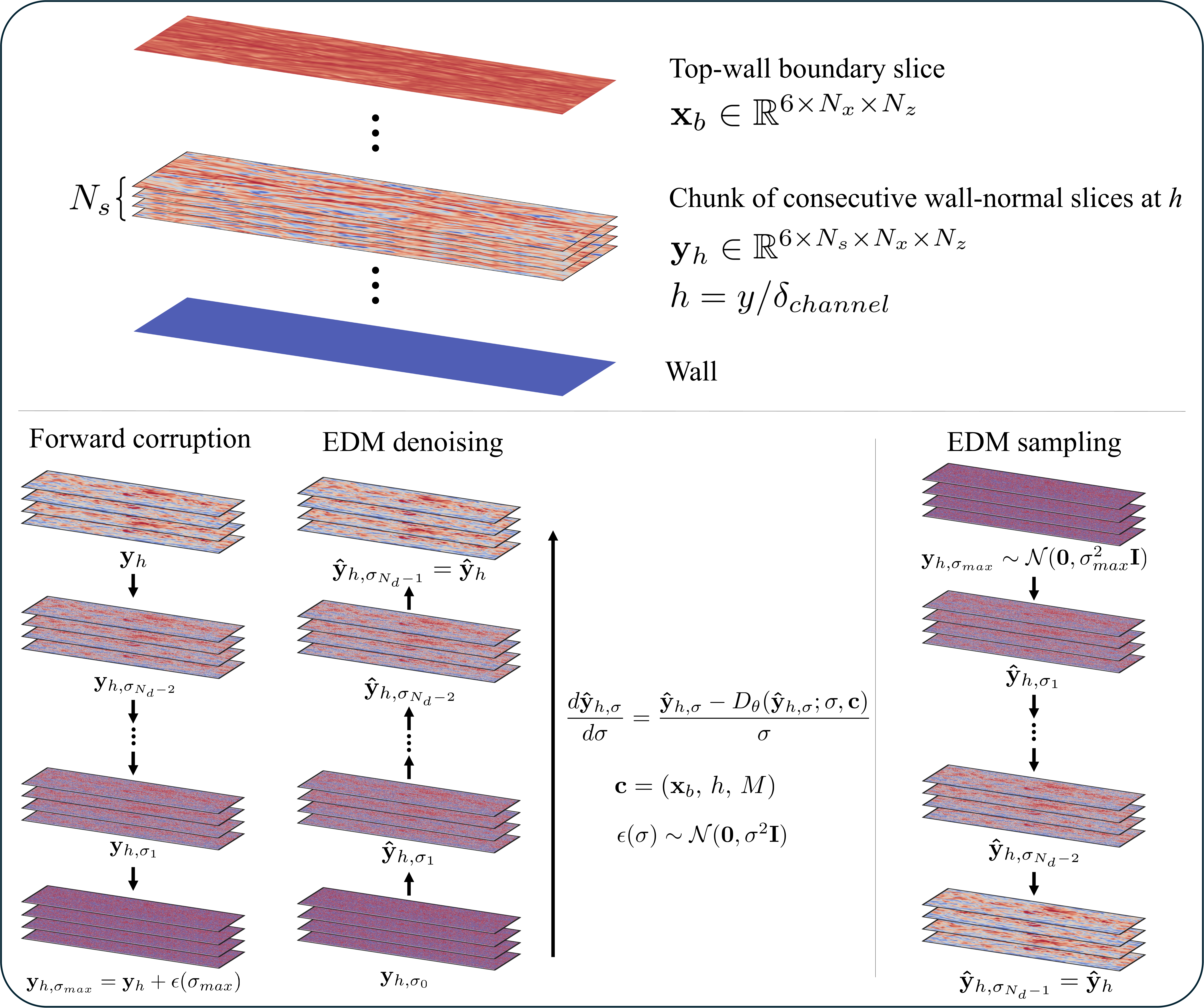}
\caption{Overview of the subdomain-wise conditional diffusion framework. Top: the top-wall boundary slice $\mathbf{x}_b$ and the target subdomain $\mathbf{y}_h$ of $N_s$ consecutive wall-normal slices starting at normalized height $h = y/\delta_{channel}$. Bottom: forward corruption, EDM denoising via the probability-flow ODE, and EDM sampling, with conditioning $\mathbf{c} = (\mathbf{x}_b, h, M)$ (see Sec.~\ref{subsec:diffusion}).}
\label{fig:flowchart}
\end{figure}

Instead, we reformulate the problem as generating a \emph{subdomain} of $N_s$ consecutive wall-normal slices starting at a specified wall-normal location within the boundary layer. Concretely, given a top-wall boundary slice $\mathbf{x}_b$, a Mach number $M$, and a target normalized wall-normal height $h$, the model generates a subdomain
\begin{equation}
    \mathbf{y}_h \in \mathbb{R}^{6 \times N_s \times N_x \times N_z}, \qquad N_s = 4,
\end{equation}
representing the six flow variables over the $N_s$ consecutive wall-normal planes whose first plane is located at $h$. The subdomain depth $N_s = 4$ was selected on the basis of preliminary experiments on a smaller subset of the dataset, balancing per-subdomain computational cost against the model's ability to capture wall-normal coherence within each subdomain; smaller values fragment the wall-normal structure excessively, while larger values increase memory requirements. Here $h$ is the normalized wall-normal coordinate,
\begin{equation}
    h = y/\delta_{channel},
\end{equation}
where $\delta_{channel}=2\delta$ is the channel height (see Sec.~\ref{sec:dataset}), so that $h \in [0, 1]$ with $h=0$ on the bottom wall and $h=1$ on the top (driven) wall. With this convention, the height conditioner $h$ denotes the lower bound of the subdomain, and the subdomain extends from $h$ toward increasing wall-normal coordinate. This subdomain-wise design keeps the per-sample cost manageable while allowing wall-normal position and Mach number to act as explicit conditioners. Section~\ref{subsec:fullvolume} describes how subdomains are composed into the full volume. Figure~\ref{fig:flowchart} summarizes the setup (top) and previews the EDM-based generation pipeline (bottom), which is detailed in Sec.~\ref{subsec:diffusion}.
 
\subsection{Conditional Diffusion Model (EDM)}
\label{subsec:diffusion}
 
For diffusion-based generative modeling, we adopt the Elucidating Diffusion Model (EDM) framework~\citep{karras2022elucidating}. In EDM, the forward process corrupts a clean sample $\mathbf{y}_h$ with Gaussian noise at level $\sigma$,
\begin{equation}
    \mathbf{y}_{h,\sigma} = \mathbf{y}_h + \boldsymbol{n}, \qquad \boldsymbol{n} \sim \mathcal{N}(\mathbf{0}, \sigma^2 \mathbf{I}),
\end{equation}
and the model learns a denoiser $D_\theta(\mathbf{y}_{h,\sigma}; \sigma, \mathbf{c})$ that predicts the clean sample from the noisy input. The conditioning vector is
\begin{equation}
    \mathbf{c} = (\mathbf{x}_b, h, M),
\end{equation}
i.e., the top-wall boundary slice, the normalized wall-normal height, and the freestream Mach number.
 
% \noindent\textbf{EDM preconditioning.} 
To ensure well-conditioned training across noise levels, the denoiser adopts the preconditioning of \citet{karras2022elucidating}:
\begin{equation}
    D_\theta(\mathbf{y}_{h,\sigma}; \sigma, \mathbf{c})
    = c_\text{skip}(\sigma)\,\mathbf{y}_{h,\sigma}
    + c_\text{out}(\sigma)\, F_\theta\!\left(c_\text{in}(\sigma)\,\mathbf{y}_{h,\sigma},\; c_\text{noise}(\sigma),\; \mathbf{c}\right),
\end{equation}
with
\begin{equation}
\begin{aligned}
c_\text{skip}(\sigma) &= \frac{\sigma_\text{data}^2}{\sigma^2 + \sigma_\text{data}^2}, &
c_\text{out}(\sigma)  &= \frac{\sigma\,\sigma_\text{data}}{\sqrt{\sigma^2 + \sigma_\text{data}^2}}, \\
c_\text{in}(\sigma)   &= \frac{1}{\sqrt{\sigma^2 + \sigma_\text{data}^2}}, &
c_\text{noise}(\sigma) &= \tfrac{1}{4}\ln\sigma.
\end{aligned}
\end{equation}
 
% \noindent\textbf{Training objective.} 
The network $F_\theta$ is trained with the weighted EDM denoising loss,
\begin{equation}
    \mathcal{L}_\text{EDM}(\theta) =
    \mathbb{E}_{(\mathbf{c},\mathbf{y}_h),\,\sigma,\,\boldsymbol{n}}
    \left[
    \lambda(\sigma)\,
    \left\| D_\theta(\mathbf{y}_{h,\sigma}; \sigma, \mathbf{c}) - \mathbf{y}_h \right\|_2^2
    \right],
    \label{eq:edm_loss}
\end{equation}
where $\lambda(\sigma) = (\sigma^2 + \sigma_\text{data}^2)/(\sigma\,\sigma_\text{data})^2$ emphasizes low-noise regimes, and the training noise level is sampled as $\ln\sigma \sim \mathcal{N}(P_\text{mean}, P_\text{std}^2)$.
 
% \noindent\textbf{Sampling.}
At inference, samples are drawn by integrating the probability-flow ODE

\begin{equation}
\dfrac{d\mathbf{\hat{y}}_{h,\sigma}}{d\sigma}=\dfrac{\mathbf{\hat{y}}_{h,\sigma}-D_{\theta}(\mathbf{\hat{y}}_{h,\sigma};\sigma,\mathbf{c})}{\sigma}
\label{eq:prob_flow_ode}
\end{equation}

from $\mathbf{y}_{h,\sigma_\text{max}} \sim \mathcal{N}(\mathbf{0}, \sigma_\text{max}^2\mathbf{I})$ down to $\sigma \to \sigma_{min}$ with the EDM deterministic sampler. Because the only stochasticity enters through the initial noise, we obtain an ensemble of plausible reconstructions by drawing multiple initial samples for the same conditioning $\mathbf{c}$, which directly yields uncertainty estimates on all downstream statistics.
 
% \noindent\textbf{Architecture and conditioning.} 
The denoiser $F_\theta$ is implemented as a SongUNet~\citep{song2020score}, a U-Net with residual blocks and multi-head self-attention. The top-wall boundary slice $\mathbf{x}_b$ is concatenated channel-wise to the scaled input $c_\text{in}(\sigma)\,\mathbf{y}_{h,\sigma}$, serving as a dense spatial conditioner. The normalized wall-normal height $h$ and freestream Mach number $M$ are projected through separate linear layers and added to the noise-level embedding:
\begin{equation}
    \mathbf{e} = \mathbf{e}_\sigma + \phi_h(h) + \phi_M(M),
\end{equation}
where $\mathbf{e}_\sigma$ is the positional embedding of $c_\text{noise}(\sigma)$, and $\phi_h, \phi_M$ are learned linear projections. The combined embedding $\mathbf{e}$ is injected into each residual block via adaptive group normalization (AdaGN), modulating feature maps as a function of noise level, wall-normal position, and flow condition. Further EDM details are provided in Appendix~\ref{app:edm}.

The denoiser operates on two-dimensional ($x, z$) feature maps. The $N_s = 4$ wall-normal slices of the subdomain and the six flow variables are stacked along the channel dimension before being fed to the network: a subdomain $\mathbf{y}_h \in \mathbb{R}^{6 \times N_s \times N_x \times N_z}$ is reshaped into a tensor of shape $(N_s \cdot 6) \times N_x \times N_z = 24 \times N_x \times N_z$, and the conditioning slice $\mathbf{x}_b \in \mathbb{R}^{6 \times N_x \times N_z}$ is concatenated along the channel dimension to give a $30$-channel input to the U-Net. The network output of shape $24 \times N_x \times N_z$ is reshaped back to $6 \times N_s \times N_x \times N_z$ to recover the multivariate subdomain. Within this representation, all spatial convolutions act on the homogeneous $(x, z)$ plane, while wall-normal coherence within the subdomain is captured implicitly through the channel-mixing operations of the residual blocks, the self-attention layers, and the group normalization.
 
\subsection{Bounded Binned Spectral-Power (BSP) Loss for Diffusion Model}
\label{subsec:bsp}
 Accurate high-wavenumber content is critical for turbulence reconstruction but is not directly penalized by the pointwise EDM loss. To explicitly encourage spectral fidelity, we augment training with a binned spectral-power (BSP) loss \citep{chakraborty2026binned} evaluated on binned one-dimensional energy spectra along the streamwise and spanwise directions. The original BSP loss takes the ratio form
\begin{equation}
    \mathcal{L}_\text{BSP}^\text{ratio}(u, v) = \frac{1}{N_k}\sum_{c=1}^{C}\sum_{i=1}^{N_k}
    \left(1 - \frac{E_u^\text{bin}(c,i) + \epsilon}{E_v^\text{bin}(c,i) + \epsilon}\right)^2,
    \label{eq:bsp_ratio}
\end{equation}
where $N_k$ is the number of wavenumber bins, $C$ is the number of channels, and $E_u^\text{bin}$, $E_v^\text{bin}$ are binned energy spectra of the generated field $u$ and target $v$.
 
Because diffusion-based models are trained across a wide range of noise levels, the ratio form of Eq.~\eqref{eq:bsp_ratio} can become numerically problematic. Although the EDM preconditioner normalizes the denoiser output $D_\theta$ to remain close to the data scale, the predicted spectrum $E_u$ computed from $D_\theta$ can still deviate substantially from the target spectrum $E_v$, particularly early in optimization when the denoiser has not yet converged and the per-bin spectral ratio $E_u / E_v$ can fluctuate widely. In these regimes, the ratio form produces noisy and potentially destabilizing gradient signals. To remedy this, we use a bounded, symmetric variant,

\begin{equation}
    \mathcal{L}_\text{BSP}(u, v) = \frac{1}{C}\frac{1}{N_k}\sum_{c=1}^{C}\sum_{i=1}^{N_k}
    \left(\frac{E_u^\text{bin}(c,i,\sigma) - E_v^\text{bin}(c,i,\sigma)}{E_u^\text{bin}(c,i,\sigma) + E_v^\text{bin}(c,i,\sigma) + \epsilon}\right)^2,
    \label{eq:bsp_bounded}
\end{equation}
which normalizes the difference by the sum rather than taking a ratio. Because $(E_u - E_v)^2 \leq (E_u + E_v)^2$ for all $E_u, E_v \geq 0$, each term is bounded to $[0,1]$ regardless of the magnitude of $E_u$ or $E_v$, yielding well-behaved gradients throughout optimization and across the entire noise schedule.
 
In practice, we use the composite objective
\begin{equation}
    \mathcal{L}(\theta) =
    \mathbb{E}_{(\mathbf{c},\mathbf{y}_h),\,\sigma,\,\boldsymbol{n}}
    \left[
    \lambda(\sigma)\,
    \left(
    \left\| D_\theta(\mathbf{y}_{h,\sigma}; \sigma, \mathbf{c}) - \mathbf{y}_h \right\|_2^2
    \;+\; \beta\,\mathcal{L}_\text{nBSP}(D_\theta, \mathbf{y}_h)
    \right)
    \right],
    \label{eq:composite_loss}
\end{equation}
where $\lambda(\sigma) = (\sigma^2 + \sigma_\text{data}^2)/(\sigma\,\sigma_\text{data})^2$ is the EDM weighting (Sec.~\ref{subsec:diffusion}) and $\beta = 10^{-3}$ is a scalar weight chosen empirically so that the two terms in the composite loss remain of comparable magnitude during training. The shared $\lambda(\sigma)$ weighting ensures that both loss terms emphasize low-noise regimes, where the denoiser prediction is closest to the clean target and the spectral content is most physically meaningful, while downweighting the high-noise regime and further mitigating any residual sensitivity of the BSP loss to early training inaccuracies.
 
\subsection{Full-Volume Reconstruction via Soft Overlap Inpainting}
\label{subsec:fullvolume}
 
The subdomain-wise formulation yields samples over $N_s$-slice windows, which must be composed into a full wall-normal stack of $N_y = 192$ planes. We consider three strategies:
 
\noindent\textbf{(i) No overlap.} Adjacent subdomains are generated independently at non-overlapping heights and concatenated. This strategy is simple but produces visible discontinuities at subdomain interfaces because the model has no mechanism to enforce continuity across subdomains.
 
\noindent\textbf{(ii) Overlap with averaging.} Adjacent subdomains share a common overlap region in wall-normal space, and the overlap slices are replaced by the arithmetic average of the two independent predictions. This removes seams but, because averaging two stochastic samples reduces the variance of small-scale content, systematically damps turbulent fluctuations in the overlapping planes.
 
\begin{figure}
\centering
\includegraphics[width=\textwidth]{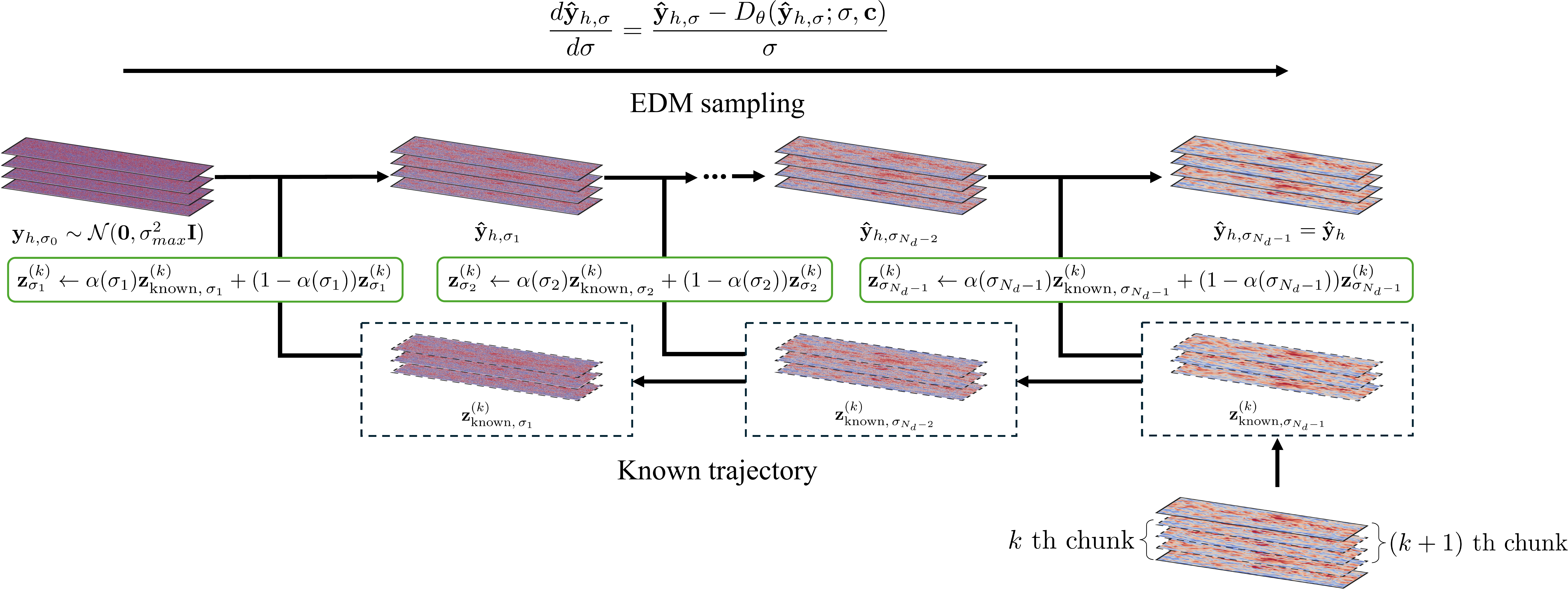}
\caption{Soft overlap inpainting for subdomain-to-subdomain composition. \emph{Top (EDM sampling):} reverse-diffusion trajectory of the $(k{+}1)$-th subdomain from pure noise. \emph{Bottom (known trajectory):} noised versions of the overlap region $\mathbf{z}^{(k)}_\text{known}$ inherited from the previously generated $k$-th subdomain.}
\label{fig:inpaint_schematic}
\end{figure}

\noindent\textbf{(iii) Soft overlap inpainting.} To preserve continuity \emph{and} statistical variability, we use a replacement-based inpainting scheme during the reverse diffusion process, inspired by RePaint~\cite{lugmayr2022repaint} and adapted to subdomain-to-subdomain composition along the wall-normal direction. The idea is illustrated in Fig.~\ref{fig:inpaint_schematic}. Consider the generation of the $(k{+}1)$-th subdomain, which overlaps the previously generated $k$-th subdomain on a set of wall-normal planes. We denote by $\mathbf{z}_\text{known}^{(k)}$ the restriction of the already generated $k$-th subdomain to this overlap region, which is treated here as a known ground-truth trajectory for the overlap.
 
During sampling of the $(k{+}1)$-th subdomain, we maintain two streams at each noise level $\sigma_t$ (Fig.~\ref{fig:inpaint_schematic}):
\begin{enumerate}
    \item A \emph{known trajectory} obtained by noising $\mathbf{z}_\text{known}^{(k)}$ forward along the EDM schedule,
    \begin{equation}
        \mathbf{z}^{(k)}_{\text{known},\sigma_t}
        = \mathbf{z}_\text{known}^{(k)} + \epsilon(\sigma_t),
        \qquad \epsilon(\sigma_t)\sim\mathcal{N}(\mathbf{0},\sigma_t^2\mathbf{I}),
    \end{equation}
    which provides a continuity anchor to the $k$-th subdomain at every noise level.
    \item The \emph{model trajectory } $\mathbf{z}^{(k)}_{\sigma_t}$, i.e., the restriction of the current reverse-diffusion sample of the $(k{+}1)$-th subdomain to the overlap region, produced by the probability-flow ODE of Sec.~\ref{subsec:diffusion}.
\end{enumerate}
At each reverse-diffusion step, after the ODE update, we replace the overlap region of the current subdomain with a soft blend of the two streams:
\begin{equation}
    \mathbf{z}^{(k)}_{\sigma_t}
    \leftarrow
    \alpha(\sigma_t)\,\mathbf{z}^{(k)}_{\text{known},\sigma_t}
    + \bigl(1-\alpha(\sigma_t)\bigr)\,\mathbf{z}^{(k)}_{\sigma_t},
    \label{eq:soft_inpaint}
\end{equation}
where $\alpha(\sigma_t)\in[0,1]$ is a blending coefficient that controls how strongly the $(k{+}1)$-th subdomain is anchored to the $k$-th subdomain in the overlap region. Early in the reverse process (large $\sigma_t$, $\alpha$ relatively large) the known trajectory dominates, enforcing inter-subdomain continuity; late in the process (small $\sigma_t$, $\alpha$ small) the model's own prediction dominates, so that fine-scale turbulent variance is not erased. In practice, we use the cosine schedule
\begin{equation}
    \alpha(\sigma_t) = \tfrac{1}{4}\!\left(1 + \cos\frac{\pi (t-1)}{N_d - 2}\right),
    \label{eq:alpha_schedule}
\end{equation}
where $t \in \{1, \ldots, N_d-1\}$ indexes the discretized noise levels along the probability-flow ODE and $N_d$ is the total number of denoising steps. This schedule is monotonically decreasing in $t$ (i.e., monotonically decreasing along the reverse-diffusion direction), with $\alpha=\tfrac{1}{2}$ at the start ($t=1$) and $\alpha=0$ at the end ($t=N_d{-}1$). Pseudo-code is provided in Appendix~\ref{app:inpainting}.
 
Figure~\ref{fig:inpaint_schematic} illustrates this mechanism: the top row is the reverse-diffusion (EDM sampling) trajectory of the $(k{+}1)$-th subdomain, from pure noise $\mathbf{y}_{h,\sigma_{\max}}\sim\mathcal{N}(\mathbf{0},\sigma_{\max}^2\mathbf{I})$ to the clean sample $\hat{\mathbf{y}}_{h,\sigma_{min}} = \hat{\mathbf{y}}_h$; the bottom row is the known trajectory obtained by noising the $k$-th subdomain's overlap region; the blending at each noise level follows Eq.~\eqref{eq:soft_inpaint} with the schedule \eqref{eq:alpha_schedule}. The two simpler strategies (no overlap and averaging) are direct baselines that require no additional illustration; the three are compared quantitatively on actual reconstructions in Sec.~\ref{subsec:training_case}.
 
\section{Results}
\label{sec:Results}

We evaluate the proposed framework on the hypersonic Couette DNS dataset of ~\citet{toki2024sub}. A single Mach-conditioned diffusion model is trained jointly on $M=6,7,8$ and evaluated at each of these three in-distribution Mach numbers. All evaluations throughout this section are performed on a held-out subset of DNS snapshots that is disjoint from the training data: at each Mach number, $80\%$ of the available snapshots are used for training and the remaining $20\%$ are withheld and used exclusively for sampling and evaluation. Throughout this section, the results are organized to isolate the two methodological axes introduced in this work. The composition strategy (Sec.~\ref{subsec:fullvolume}) and the bounded BSP loss (Sec.~\ref{subsec:bsp}) operate at different stages of the pipeline; the former during inference and the latter during training, so their effects can be examined independently. For the composition ablation, we fix the training loss to EDM+BSP, since the bounded BSP loss is designed to improve spectral fidelity independently of how subdomains are assembled; its own effect on spectral statistics is isolated in a dedicated comparison immediately afterward. Section~\ref{subsec:training_case} reports results at the representative Mach number $M=6$, beginning with instantaneous fields and the composition-strategy comparison, followed by the loss ablation through spectral and statistical diagnostics, and concluding with profile uncertainty bands, spatial correlations, and uncertainty maps. Section~\ref{subsec:multimach} reports the same statistics at $M=7$ and $M=8$, documenting the performance of the single unified model across all training Mach numbers, and examines the Trettel-Larsson transformation \citep{trettel2016mean} as a compressibility-consistency diagnostic of the learned Mach conditioning.

\subsection{Representative-Mach Results ($M=6$)}
\label{subsec:training_case}

\noindent\textbf{Statistical definitions.}
Throughout this section, statistics are computed from the reconstructed full volumes as follows. Mean velocity and temperature profiles use the Favre (density-weighted) average,
\begin{equation}
    \tilde{q} = \frac{\langle\rho\,q\rangle}{\langle\rho\rangle},
\end{equation}
where $\langle\cdot\rangle$ denotes an average over the ensemble, streamwise, and spanwise directions. The mean density profile uses the Reynolds (unweighted) average $\langle\rho\rangle$. Velocity and temperature fluctuations are defined relative to the Favre mean, $q'' = q - \tilde{q}$, while density fluctuations use the Reynolds decomposition, $\rho' = \rho - \langle\rho\rangle$. The RMS profile for velocity is calculated as $\sqrt{\langle\rho\rangle\widetilde{u_i''u_i''}/\tau_w}$, where $\tau_w = \mu_w (\mathrm{d}\tilde{u} / \mathrm{d}y)|_w$ is the wall shear stress evaluated at the bottom wall using Sutherland's law for the viscosity. For the temperature and density RMS profiles,  $\sqrt{\widetilde{T''^2}}/\widetilde{T}$ and $\sqrt{\langle\rho'^2\rangle}/\langle\rho\rangle$ are used, respectively. The Reynolds shear stress is reported as $-\langle\rho u'' v''\rangle / \tau_w$. Two-point correlations are computed via FFT in the homogeneous streamwise and spanwise directions and normalized to unity at zero separation. The semi-local wall unit is defined as $y^* \equiv y\sqrt{\langle\rho\rangle|\tau_w|}/\langle\mu\rangle$, where $\langle\mu\rangle$ is evaluated through Sutherland's law from the local mean temperature~\citep{huang1995compressible}.

\begin{figure}
\centering
\includegraphics[width=\textwidth]{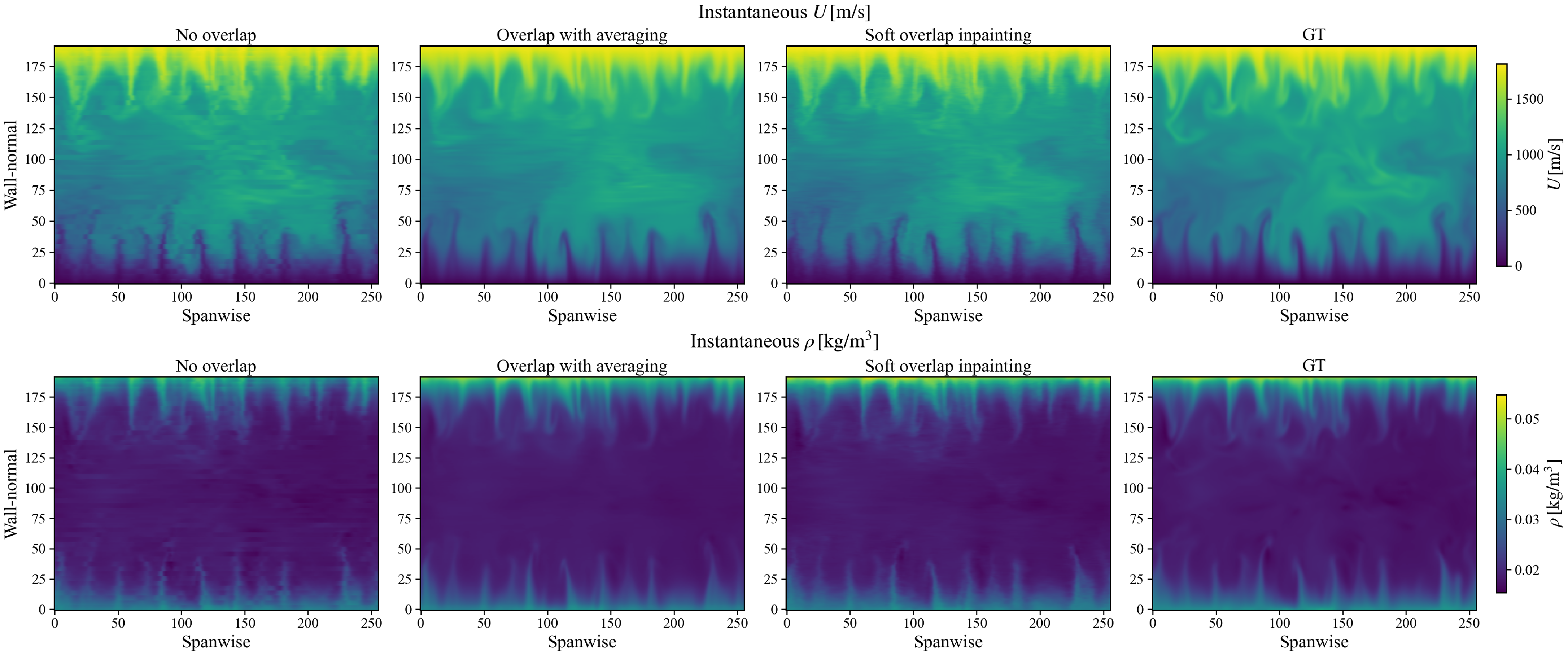}
\caption{Full-volume wall-normal-spanwise slices of instantaneous streamwise velocity $U$ (top row) and pressure $P$ (bottom row) at $M=6$. From left to right: no overlap, overlap with averaging, soft overlap inpainting, and DNS ground truth (GT). The no-overlap reconstruction exhibits visible horizontal seams at subdomain interfaces, while both overlap-averaging and soft overlap inpainting produce visually continuous full-volume fields.}
\label{fig:slice_M6_volume}
\end{figure}

\noindent\textbf{Composition-strategy comparison.}
We first compare the three composition strategies introduced in Sec.~\ref{subsec:fullvolume}. Since the bounded BSP loss is designed to improve spectral fidelity independently of how subdomains are assembled (Sec.~\ref{subsec:bsp}), we fix the training loss to EDM+BSP and vary only the composition strategy; the effect of the BSP loss itself is isolated in the spectral comparison that follows. Figure~\ref{fig:slice_M6_volume} shows full-volume wall-normal-spanwise slices of instantaneous streamwise velocity $U$ and pressure $P$ under the three strategies, alongside the DNS ground truth. The no-overlap reconstruction exhibits visible horizontal seams at subdomain interfaces, where independently sampled subdomains fail to align across the boundary. Both overlap averaging and soft overlap inpainting remove these seams and produce visually more continuous full-volume fields than the no-overlap reconstruction. 

\begin{figure}
\centering
\includegraphics[width=\textwidth]{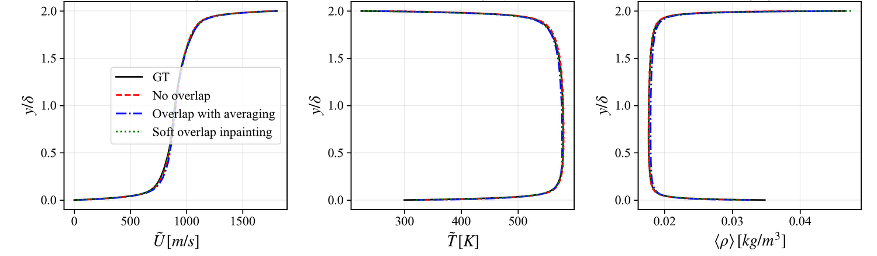}
\caption{Mean profiles of streamwise velocity (left), temperature (middle), and density (right) at $M=6$ for the three composition strategies compared against DNS (GT). First-order statistics are recovered accurately by all three strategies, consistent with their relative insensitivity to local fluctuation amplitudes.}
\label{fig:mean_profiles_M6_volume}
\end{figure}

\begin{figure}
\centering
\includegraphics[width=\textwidth]{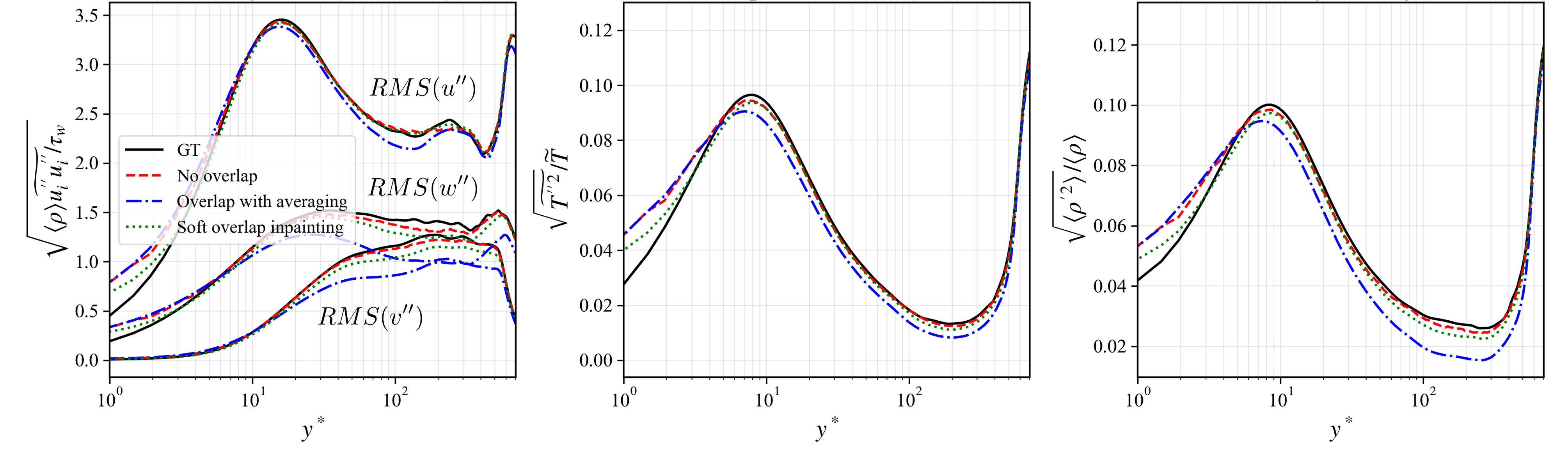}
\caption{RMS profiles of velocity components $u''$, $v''$, $w''$ (left), temperature (middle), and density (right) at $M=6$ for the three composition strategies compared against DNS (GT).  Overlap averaging systematically damps the fluctuation intensities throughout the boundary layer, while no overlap and soft overlap inpainting recover the DNS RMS levels closely across all three velocity components and the thermophysical variables.}
\label{fig:RMS_profiles_M6_volume}
\end{figure}

\begin{figure}
\centering
\includegraphics[width=\textwidth]{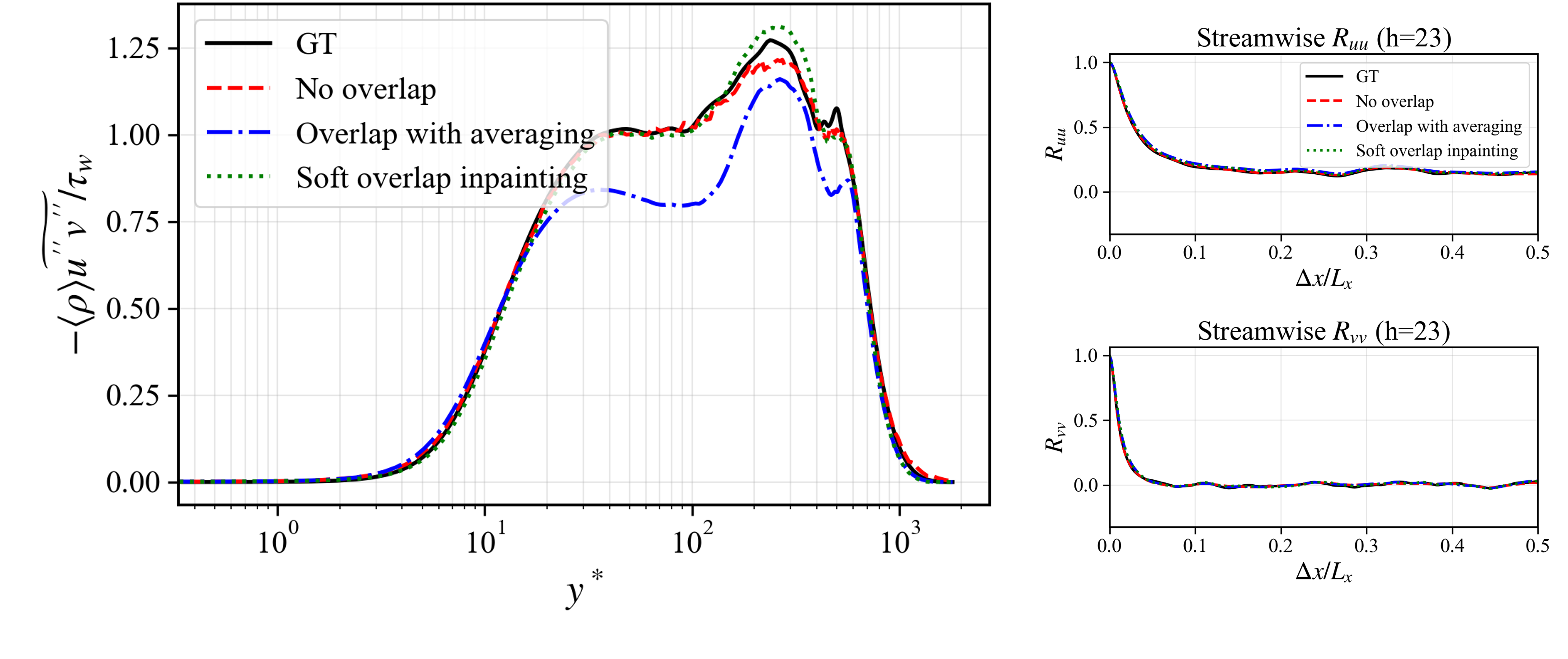}
\caption{Reynolds shear stress $-\langle \rho u'' v''\rangle / \tau_w$ (left) and streamwise and spanwise two-point correlations $R_{uu}$ at $y/\delta = 0.0352$ (right) at $M=6$ for the three composition strategies compared against DNS (GT). Overlap averaging substantially underestimates the Reynolds-shear peak, whereas soft overlap inpainting matches the DNS profile closely; the two-point correlations are reproduced comparably well by all three strategies and capture the expected streamwise-spanwise anisotropy.}
\label{fig:Re_shear_correlation_M6_volume}
\end{figure}

The effect of the composition strategy on wall-normal statistics is quantified in Figs.~\ref{fig:mean_profiles_M6_volume}-\ref{fig:Re_shear_correlation_M6_volume}. Mean profiles of streamwise velocity, temperature, and density (Fig.~\ref{fig:mean_profiles_M6_volume}) are recovered accurately by all three strategies, which is expected, since first-order statistics are largely insensitive to local fluctuation amplitudes. Higher-order statistics, however, reveal the differences between the strategies. The RMS profiles of velocity components, temperature, and density (Fig.~\ref{fig:RMS_profiles_M6_volume}) show that overlap averaging systematically underestimates the fluctuation intensities throughout the boundary layer. The no-overlap and soft-inpainting strategies both recover the RMS levels of DNS closely across all three components, in contrast to the systematic underestimation of overlap averaging. The Reynolds shear stress and two-point correlations (Fig.~\ref{fig:Re_shear_correlation_M6_volume}) confirm and refine this picture. The Reynolds shear stress profile (left) is most sensitive to the composition strategy: overlap averaging substantially underestimates the Reynolds shear stress peak, consistent with the systematic damping of $u''$ and $v''$ amplitudes seen in Fig.~\ref{fig:RMS_profiles_M6_volume}, whereas soft overlap inpainting matches the DNS profile more closely. Meanwhile, the two-point correlations $R_{uu}$ (right) are recovered comparably well by all three strategies. In particular, all three reproduce the well-known anisotropy of wall-bounded turbulence, in which the spanwise correlation decays much more rapidly than the streamwise correlation, reflecting the elongated streamwise streaks. The insensitivity of the correlations to the composition strategy is expected, since two-point correlations are normalized by the local variance and therefore probe the spatial scales of the structures rather than their absolute amplitudes.

\begin{figure}
\centering
\includegraphics[width=\textwidth]{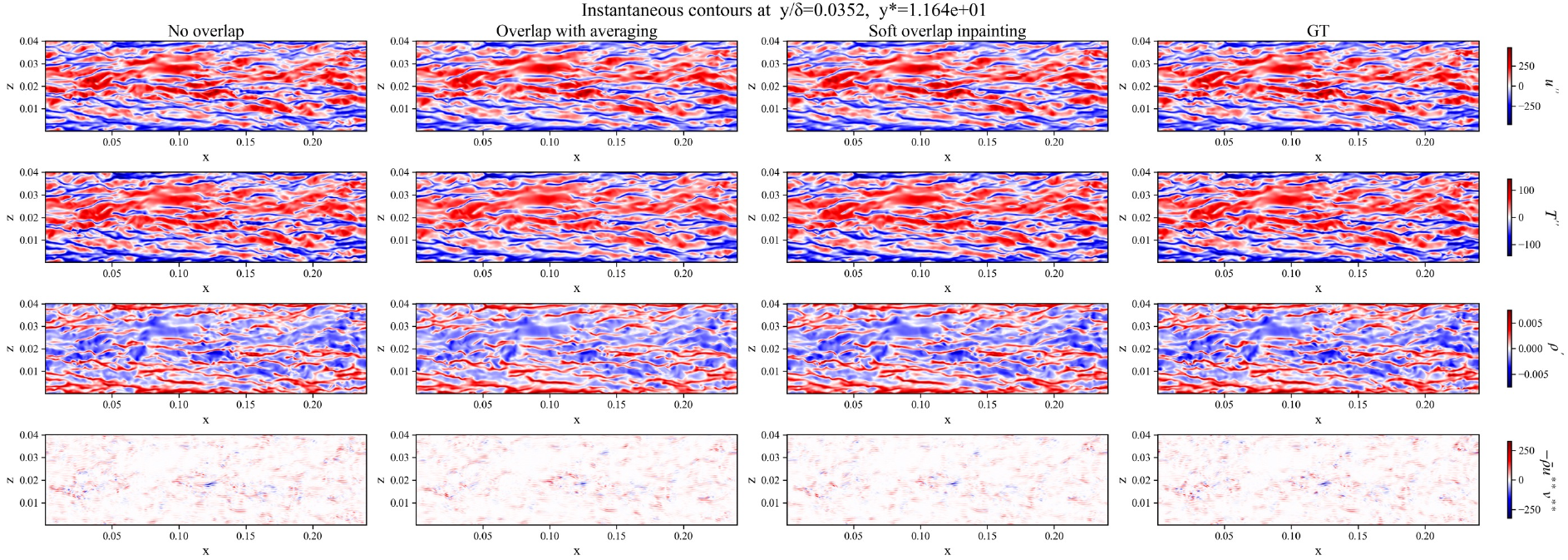}
\caption{Instantaneous wall-parallel contours at $y/\delta = 0.0352$ ($y^* \approx 11.6$), $M=6$. From left to right: no overlap, overlap with averaging, soft overlap inpainting, and DNS ground truth (GT). Rows from top to bottom: Favre velocity fluctuation $u''$, Favre temperature fluctuation $T''$, Reynolds density fluctuation $\rho'$, and the second-order residual-velocity contribution to the SFS shear stress $-\bar{\rho}\,u^{**}v^{**}$ obtained from a sharp spectral filter with cut-off wavenumbers $(k_x, k_z) = (32, 16)$. All three composition
strategies reproduce the dominant near-wall streaky structure in $u''$,
$T''$, and $\rho'$.}
\label{fig:contour_M6_near_wall_volume}
\end{figure}

\begin{figure}
\centering
\includegraphics[width=\textwidth]{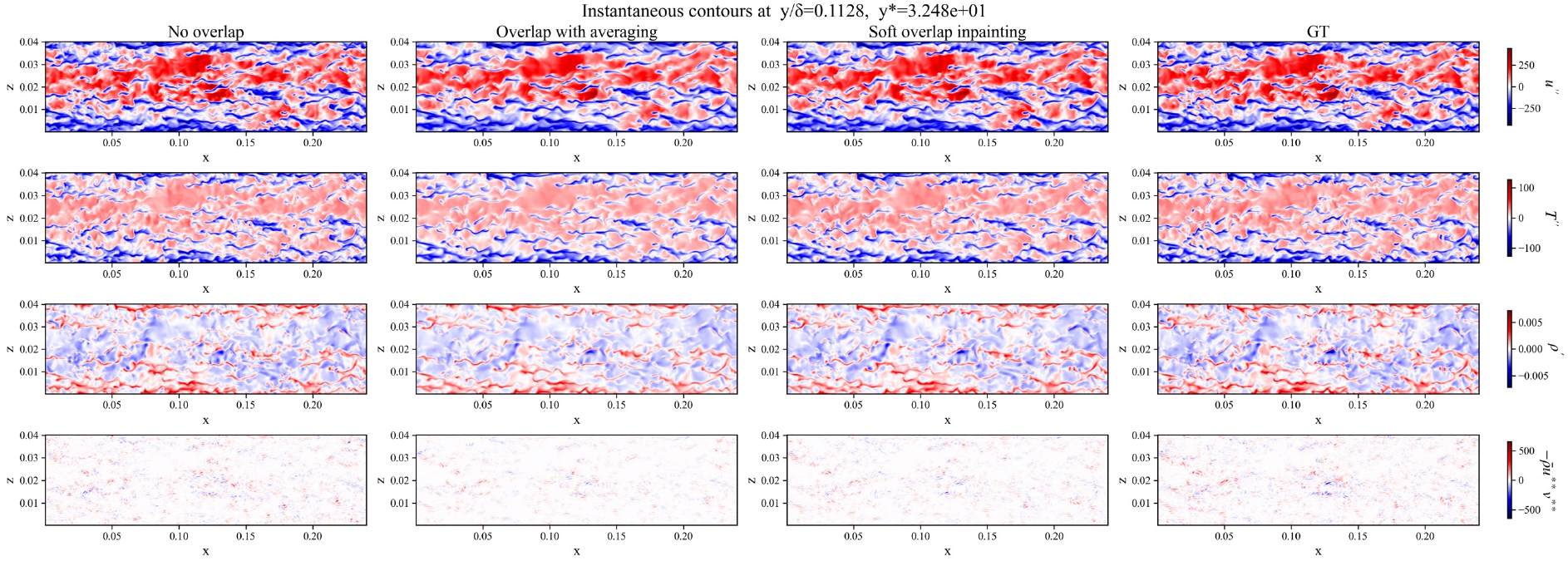}
\caption{Instantaneous wall-parallel contours at $y/\delta = 0.1128$ ($y^* \approx 32.5$), $M=6$. From left to right: no overlap, overlap with averaging, soft overlap inpainting, and DNS ground truth (GT). Rows from top to bottom: Favre velocity fluctuation $u''$, Favre temperature fluctuation $T''$, Reynolds density fluctuation $\rho'$, and the second-order residual-velocity contribution to the SFS shear stress $-\bar{\rho}\,u^{**}v^{**}$ obtained from a sharp spectral filter with cut-off wavenumbers $(k_x, k_z) = (32, 16)$. At this higher wall-normal
location the streaky structures broaden and the velocity-temperature similarity weakens, both of which are captured by all three strategies; the small-scale SFS shear-stress contribution (bottom row) separates the strategies, with soft overlap inpainting reproducing the DNS amplitudes and overlap averaging producing a weaker SFS pattern.}
\label{fig:contour_M6_near_wall_2_volume}
\end{figure}

The instantaneous wall-parallel structure of the reconstructed fluctuation fields is compared against DNS in Figs.~\ref{fig:contour_M6_near_wall_volume} and~\ref{fig:contour_M6_near_wall_2_volume} at two near-wall locations. Each figure shows four rows: the Favre velocity fluctuation $u''$, the Favre temperature fluctuation $T''$, the Reynolds density fluctuation $\rho'$, and the dominant second-order residual-velocity contribution to the sub-filter-scale (SFS) shear stress, $-\bar{\rho}\,u^{**}v^{**}$, where $\bar{(\cdot)}$ denotes a sharp spectral filter retaining only modes with $|k_x| \leq 32$ and $|k_z| \leq 16$, $\check{q} = \overline{\rho q}/\bar{\rho}$ is the corresponding filtered Favre quantity, and $q^{**} = q - \check{q}$ is the sub-filter Favre fluctuation~\citep{toki2024sub}. The first three rows probe whether each composition strategy reproduces the large-scale near-wall structure of the flow, while the fourth row probes the small-scale momentum transport that is most sensitive to averaging-induced damping.

At the lower wall-normal location (Fig.~\ref{fig:contour_M6_near_wall_volume}), all three composition strategies reproduce the dominant near-wall structure of the flow. The elongated streamwise streaks in $u''$, the closely correlated thermal streaks in $T''$, and the corresponding density fluctuations in $\rho'$ are all consistent with DNS, including the well-known correlation between high-speed/high-temperature/low-density streaks and low-speed/low-temperature/high-density streaks~\citep{pirozzoli2011turbulence, toki2021investigation, toki2024sub}. At the higher wall-normal location (Fig.~\ref{fig:contour_M6_near_wall_2_volume}), the same correlations weaken, where streamwise streaks broaden in spanwise spacing and the velocity-temperature similarity becomes less pronounced, and these features are also captured by all three strategies. The differences between strategies emerge in the SFS shear-stress contribution $-\bar{\rho}\,u^{**}v^{**}$ (bottom row): soft overlap inpainting reproduces the highly localized patchy structures of DNS with comparable amplitudes, whereas overlap averaging produces a weaker SFS pattern, consistent with the systematic variance damping observed in the RMS profiles of Fig.~\ref{fig:RMS_profiles_M6_volume}. Based on the above results, soft overlap inpainting is selected as the composition strategy: it eliminates the subdomain-interface seams of the no-overlap reconstruction (Fig.~\ref{fig:slice_M6_volume}) while avoiding the systematic amplitude damping introduced by overlap averaging (Figs.~\ref{fig:RMS_profiles_M6_volume}, \ref{fig:Re_shear_correlation_M6_volume}). All subsequent results use soft overlap inpainting.

\begin{figure}
\centering
\includegraphics[width=\textwidth]{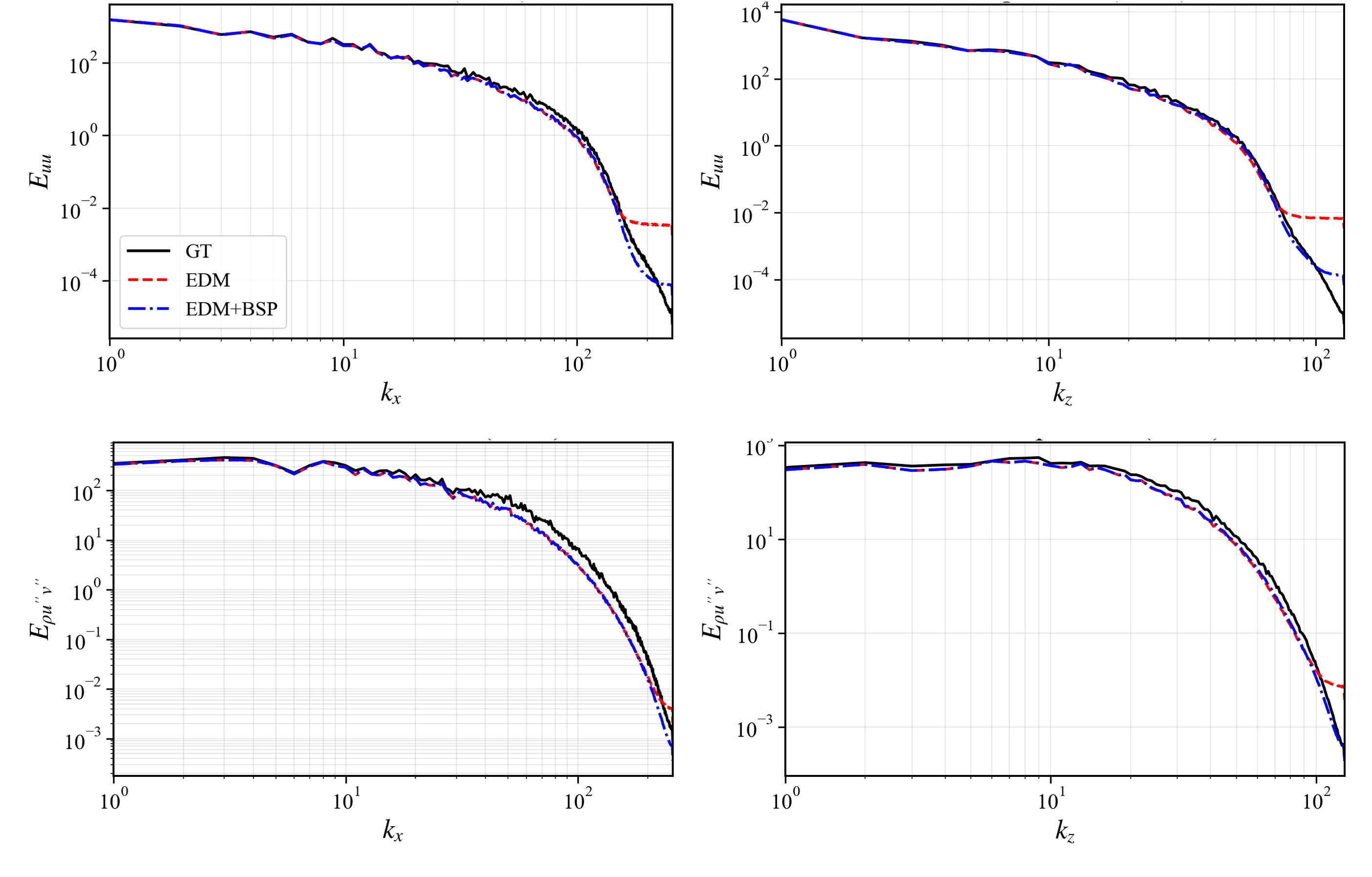}
\caption{One-dimensional energy spectra of $u''$ (top row) and Reynolds shear stress (bottom row) in the streamwise (left) and spanwise (right) directions at $y/\delta = 0.1128$, $M=6$: DNS (GT), EDM, and EDM+BSP, all using soft overlap inpainting. The bounded BSP loss suppresses the high-wavenumber excess of the baseline EDM and brings the high-wavenumber tail into close agreement with DNS in both directions, without degrading the low-wavenumber content.}
\label{fig:spectra_M6}
\end{figure}

\noindent\textbf{Loss comparison: one-dimensional energy spectra and Reynolds-stress spectra.}
Having established soft overlap inpainting as the preferred composition strategy, we now fix it and compare the baseline EDM loss against the EDM+BSP variant to isolate the effect of the bounded spectral-binning loss. The one-dimensional streamwise energy spectrum of $u''$ $E_{uu}(k_x)$ at a wall-normal location $h$ is computed from the Favre velocity fluctuation field $u''(x,z)$ on that plane by performing a one-dimensional FFT along the streamwise direction at each spanwise location, taking the squared magnitude, and averaging over the ensemble of generated samples and over the homogeneous spanwise direction. The spanwise spectrum $E_{uu}(k_z)$ is computed analogously by FFT in the spanwise direction with averaging over the streamwise direction. The one-dimensional spectrum of the Reynolds shear stress is computed in the same way but applied to the instantaneous momentum-flux field $\rho u'' v''$, where $\rho$ is the full instantaneous density (not its mean), yielding $E_{\rho u''v''}(k_x)$ and $E_{\rho u''v''}(k_z)$ in the streamwise and spanwise directions, respectively. Figure~\ref{fig:spectra_M6} shows streamwise and spanwise one-dimensional energy spectra and Reynolds-stress spectra at a representative wall-normal location ($y/\delta = 0.1128$). The baseline EDM captures the low- to mid-wavenumber energy but displays a mild high-wavenumber excess. The EDM+BSP variant sharply suppresses this excess and brings the high-wavenumber tail into close agreement with DNS, without degrading the low-wavenumber behavior, consistent with the design of the bounded BSP loss of Eq.~\eqref{eq:bsp_bounded}. The same qualitative improvement holds at other wall-normal locations and for the Reynolds-stress spectra.

\begin{figure}
\centering
\includegraphics[width=.95\textwidth]{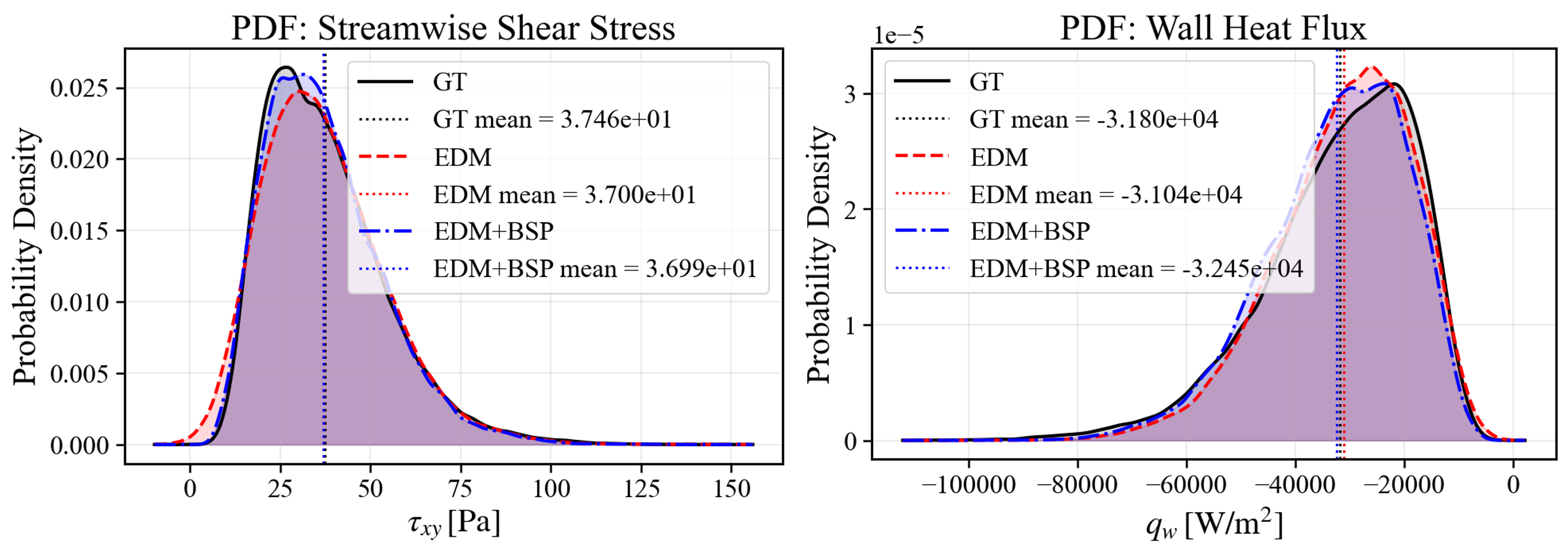}
\caption{PDFs of streamwise shear stress $\tau_{xy}$ (left) and wall heat flux $q_w$ (right) at $M=6$: DNS (GT), EDM, and EDM+BSP. Mean values for each distribution are reported in the legend. Both EDM and EDM+BSP reproduce the DNS distributions, with ensemble mean values in close quantitative agreement.}
\label{fig:tau_xy_q_bot_M6}
\end{figure}

Although the framework is trained to reconstruct interior boundary-layer fields, wall shear stress $\tau_w$ and wall heat flux $q_w$ can be recovered directly from the generated near-wall velocity and temperature gradients. Figure~\ref{fig:tau_xy_q_bot_M6} compares PDFs of streamwise wall shear stress $\tau_{xy}$ and wall heat flux $q_w$ at $M=6$. Both EDM and EDM+BSP reproduce the DNS distributions closely, with the ensemble mean values reported in the legend confirming quantitative agreement.

% \noindent\textbf{Profile uncertainty quantification.}
Because we use the EDM deterministic sampler, the only stochasticity in the sampling process enters through the initial noise. Drawing multiple samples for the same conditioning $\mathbf{c}$ therefore yields an ensemble whose spread directly quantifies the conditional posterior over $\mathbf{y}_h$ given $\mathbf{c}$. 

\begin{figure}
\centering
\includegraphics[width=\textwidth]{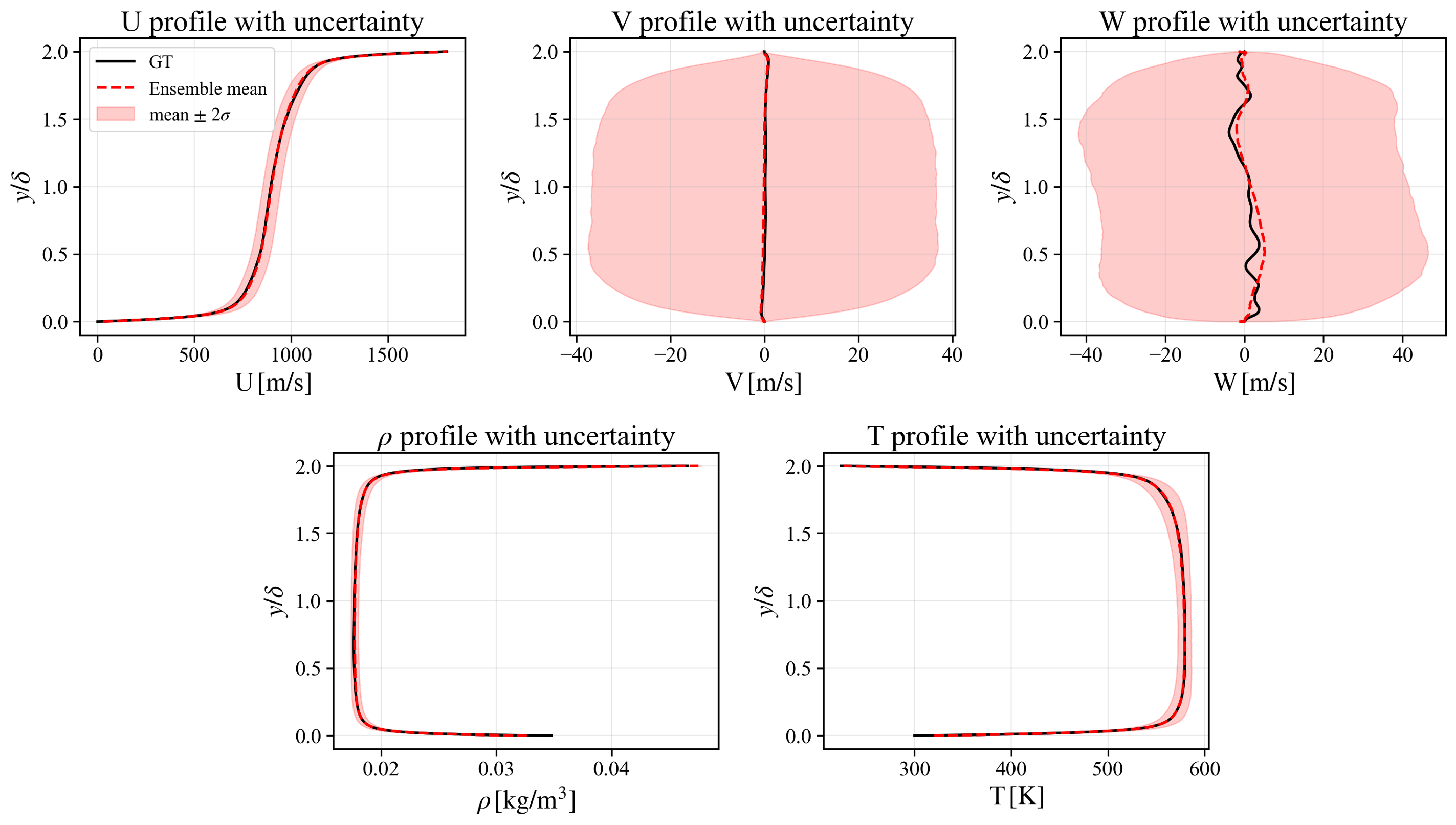}
\caption{Ensemble mean profiles (dashed red) and $\pm 2\sigma$ uncertainty bands (shaded) for $U$, $V$, $W$, $\rho$, and $T$ at $M=6$, compared against DNS. Tight uncertainty bands enclose the DNS for $U$, $\rho$, and $T$, while the fluctuation-dominated $V$ and $W$ components correctly exhibit substantially wider bands, reflecting the inherently broader conditional posterior of components with near-zero means.}
\label{fig:mean_profiles_M6_uq}
\end{figure}

Figure~\ref{fig:mean_profiles_M6_uq} shows ensemble mean profiles and $\pm 2\sigma$ uncertainty bands for $U$, $V$, $W$, $\rho$, and $T$ at $M=6$. The streamwise velocity, density, and temperature profiles are recovered with tight uncertainty bands that enclose the DNS profile, while the wall-normal and spanwise velocity components exhibit substantially wider bands. This contrast is physically consistent: $V$ and $W$ have near-zero means and are dominated by fluctuations, so the conditional posterior is inherently broader for these components, and the model correctly reflects this.
 
\begin{figure}
\centering
\includegraphics[width=\textwidth]{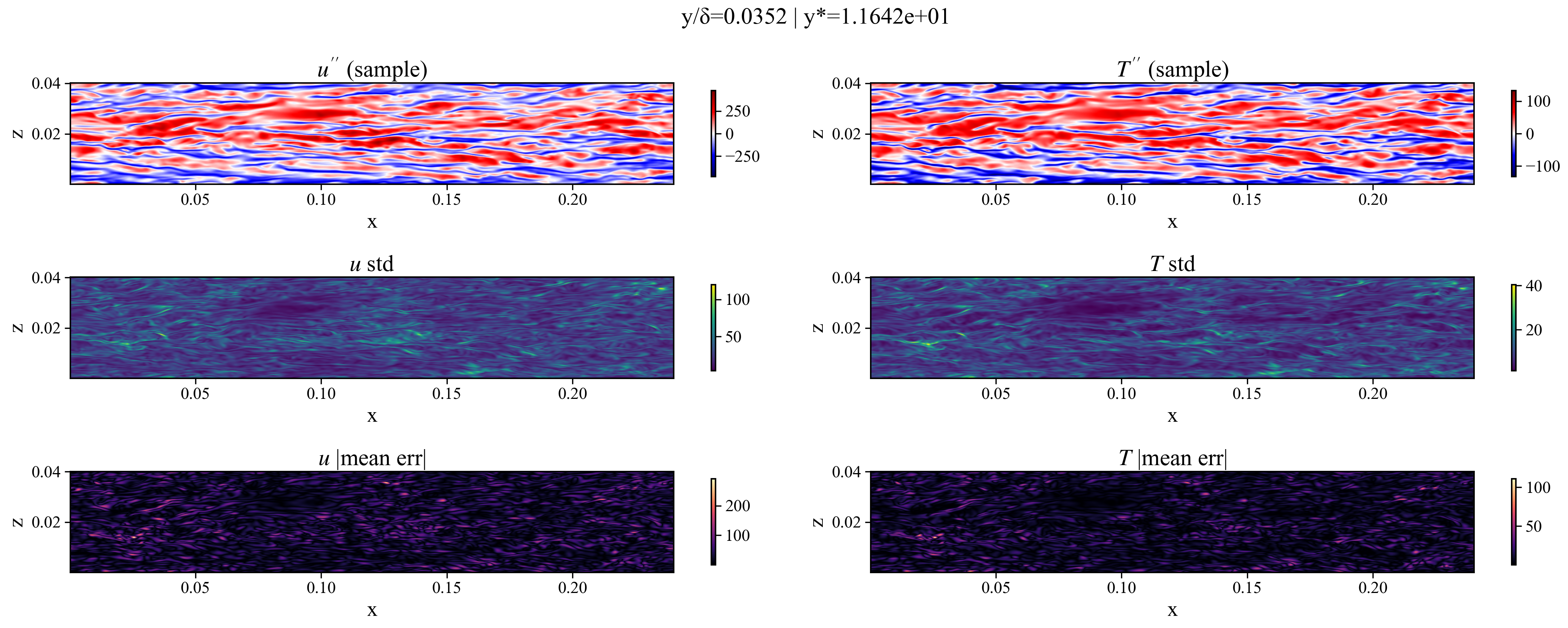}
\caption{Spatial uncertainty analysis at $y/\delta = 0.0352$ ($y^* \approx 11.6$), $M=6$. Top: instantaneous fluctuation fields $u''$ (left) and $T''$ (right) from a single generated sample. Middle: ensemble predictive standard deviation and Bottom: absolute mean error for the streamwise velocity and temperature. Predictive uncertainty concentrates along inter-streak boundaries and spatially correlates with the absolute mean error, indicating that the ensemble spread serves as a reliable indicator of local reconstruction fidelity.}
\label{fig:UQcontour_M6_near_wall}
\end{figure}

\begin{figure}
\centering
\includegraphics[width=\textwidth]{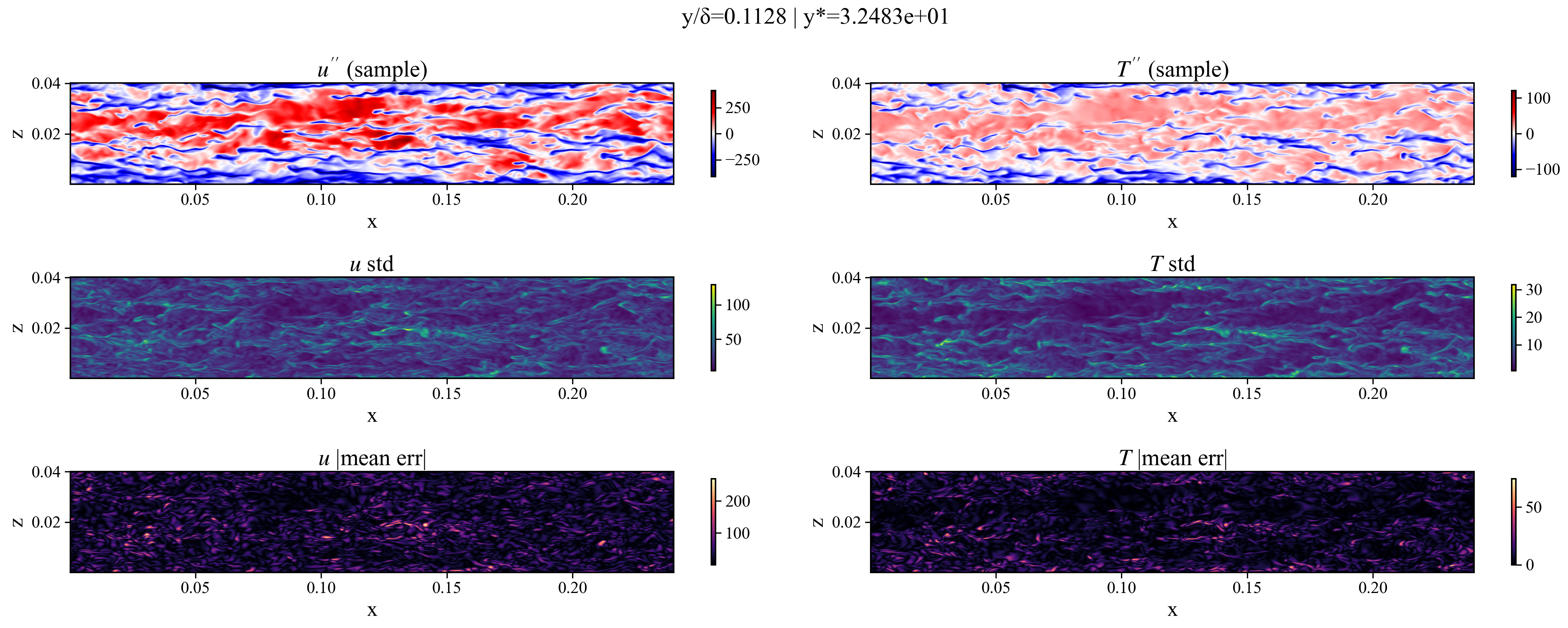}
\caption{Spatial uncertainty analysis at $y/\delta = 0.1128$ ($y^* \approx 32.5$), $M=6$. Top: instantaneous fluctuation fields $u''$ (left) and $T''$ (right) from a single generated sample. Middle: ensemble predictive standard deviation and Bottom: absolute mean error for the streamwise velocity and temperature. At this higher wall-normal location the streaky structures broaden, and the spatial correlation between predictive standard deviation and absolute mean error persists.}
\label{fig:UQcontour_M6_near_wall_2}
\end{figure}

% \noindent\textbf{Spatial uncertainty maps.}
To examine the spatial distribution of predictive uncertainty, Fig.~\ref{fig:UQcontour_M6_near_wall} displays, at a near-wall plane ($y/\delta = 0.0352$, $y^* \approx 11.6$), instantaneous fluctuation fields $u''$ and $T''$ from a single generated sample (top row), the ensemble predictive standard deviation (middle row), and the absolute mean error (bottom row) for the streamwise velocity. The predictive standard deviation of $u$ exhibits coherent spatial structure that mirrors the streaky structure of the velocity fluctuation field: elevated uncertainty concentrates along inter-streak boundaries where spanwise gradients of $u''$ are strongest, while remaining lower within streak cores. This suggests that the model is more confident about the large-scale streak structure, which is partially constrained by the top-wall boundary conditioning, than about the precise lateral positioning and fine-scale modulation of streak boundaries. Importantly, the spatial distribution of predictive standard deviation closely mirrors that of the absolute mean error, indicating that the ensemble spread serves as a reliable indicator of local reconstruction fidelity. The same patterns are observed for temperature. Figure~\ref{fig:UQcontour_M6_near_wall_2} shows the same analysis at a higher wall-normal location ($y/\delta = 0.1128$, $y^* \approx 32.5$): the streaky structures broaden, the overall predictive uncertainty decreases relative to the near-wall plane, and the spatial correlation between predictive standard deviation and absolute mean error persists.
 
\subsection{Multi-Mach Performance and Learned Compressibility Scaling}
\label{subsec:multimach}
 
The model evaluated throughout this section is a \emph{single} diffusion model trained jointly across $M=6,7,8$, with Mach number supplied through the embedding $\phi_M$ (Sec.~\ref{subsec:diffusion}). In Sec.~\ref{subsec:training_case}, we report its performance at the representative Mach number $M=6$; here we report the same statistics at the other two training Mach numbers, $M=7$ and $M=8$, to document the in-distribution performance of the unified model. We then use the Trettel-Larsson transformation as a compressibility-consistency diagnostic for the learned Mach conditioning.

\begin{figure}
\centering
\includegraphics[width=\textwidth]{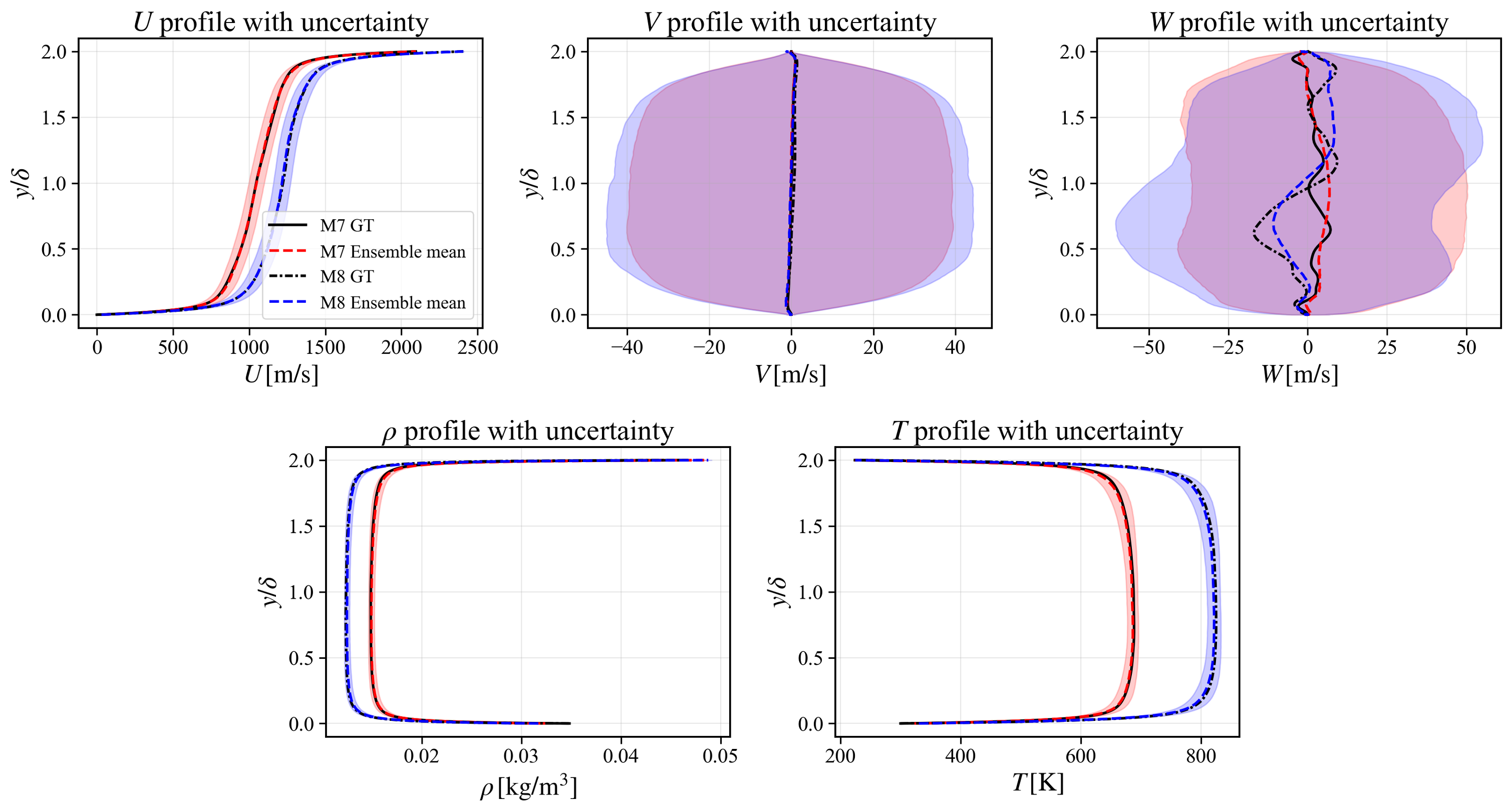}
\caption{Ensemble mean profiles (dashed) and $\pm 2\sigma$ uncertainty bands (shaded) of $U$, $V$, $W$, $\rho$, and $T$ at $M=7$ and $M=8$: generated (EDM+BSP with soft overlap inpainting) versus DNS (solid). At both Mach numbers, the ensemble mean closely tracks the DNS profile with tight bands for $U$, $\rho$, and $T$ and substantially wider bands for $V$ and $W$, in the same pattern observed at $M=6$.}
\label{fig:mean_profiles_M78_uq}
\end{figure}

Figure~\ref{fig:mean_profiles_M78_uq} shows ensemble mean profiles and $\pm 2\sigma$ uncertainty bands for $U$, $V$, $W$, $\rho$, and $T$ at $M=7$ and $M=8$, in the same format as Fig.~\ref{fig:mean_profiles_M6_uq}. At both Mach numbers, the ensemble mean closely tracks the DNS profile across the entire channel, with tight uncertainty bands for $U$, $\rho$, and $T$ and substantially wider bands for the fluctuation-dominated $V$ and $W$ components. The qualitative behavior is identical to that observed at $M=6$, demonstrating that the single jointly-trained model recovers first-order statistics with consistent uncertainty across all three training Mach numbers.

\begin{figure}
\centering
\includegraphics[width=\textwidth]{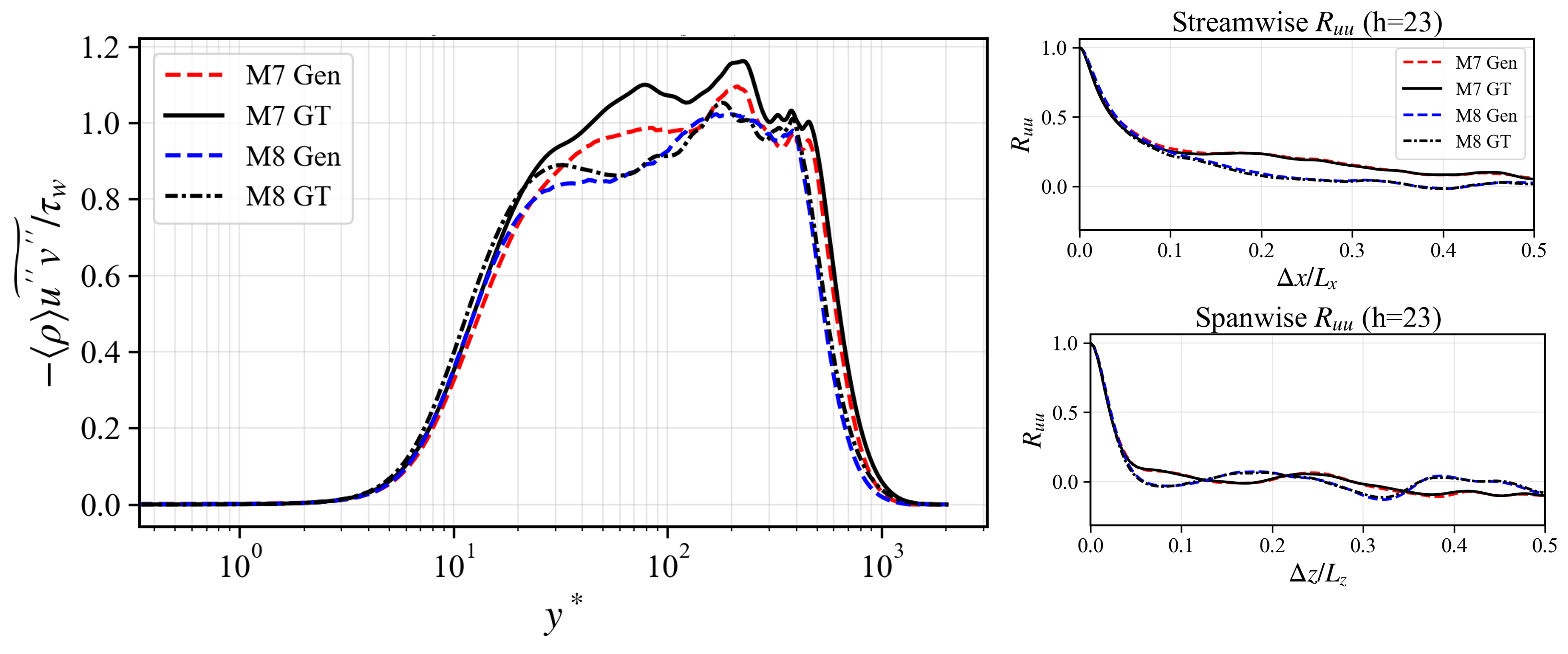}
\caption{Reynolds shear stress (left) and streamwise and spanwise two-point correlations at $y/\delta = 0.0352 $ (right)  for $M=7$ and $M=8$: generated versus DNS. The Reynolds-shear peak structure and the streamwise-spanwise correlation anisotropy are reproduced at both Mach numbers.}
\label{fig:Re_shear_correlation_M78}
\end{figure}

Figure~\ref{fig:Re_shear_correlation_M78} compares Reynolds shear stress and streamwise and spanwise two-point correlations against DNS at both Mach numbers. The Reynolds shear stress profiles are recovered, including the characteristic peak structure, and the two-point correlations exhibit two physically meaningful trends. First, at both Mach numbers, the spanwise correlation decays substantially faster than the streamwise correlation, reflecting the elongated streamwise streaks, a directional anisotropy that is well known in wall-bounded turbulence. Second, the streamwise correlation at large separations is noticeably lower at $M=8$ than at $M=7$, consistent with the higher friction Reynolds number $Re_\tau$ at $M=8$~\citep{toki2024sub}, which makes the energy-containing eddies smaller relative to the streamwise domain length and hence more rapidly decorrelated within it. Both trends are reproduced by the generative model, demonstrating that the Mach conditioning captures the corresponding variations in the spatial structure of wall-bounded turbulence, in addition to the mean profile compressibility scaling discussed below.

\begin{figure}
\centering
\includegraphics[width=\textwidth]{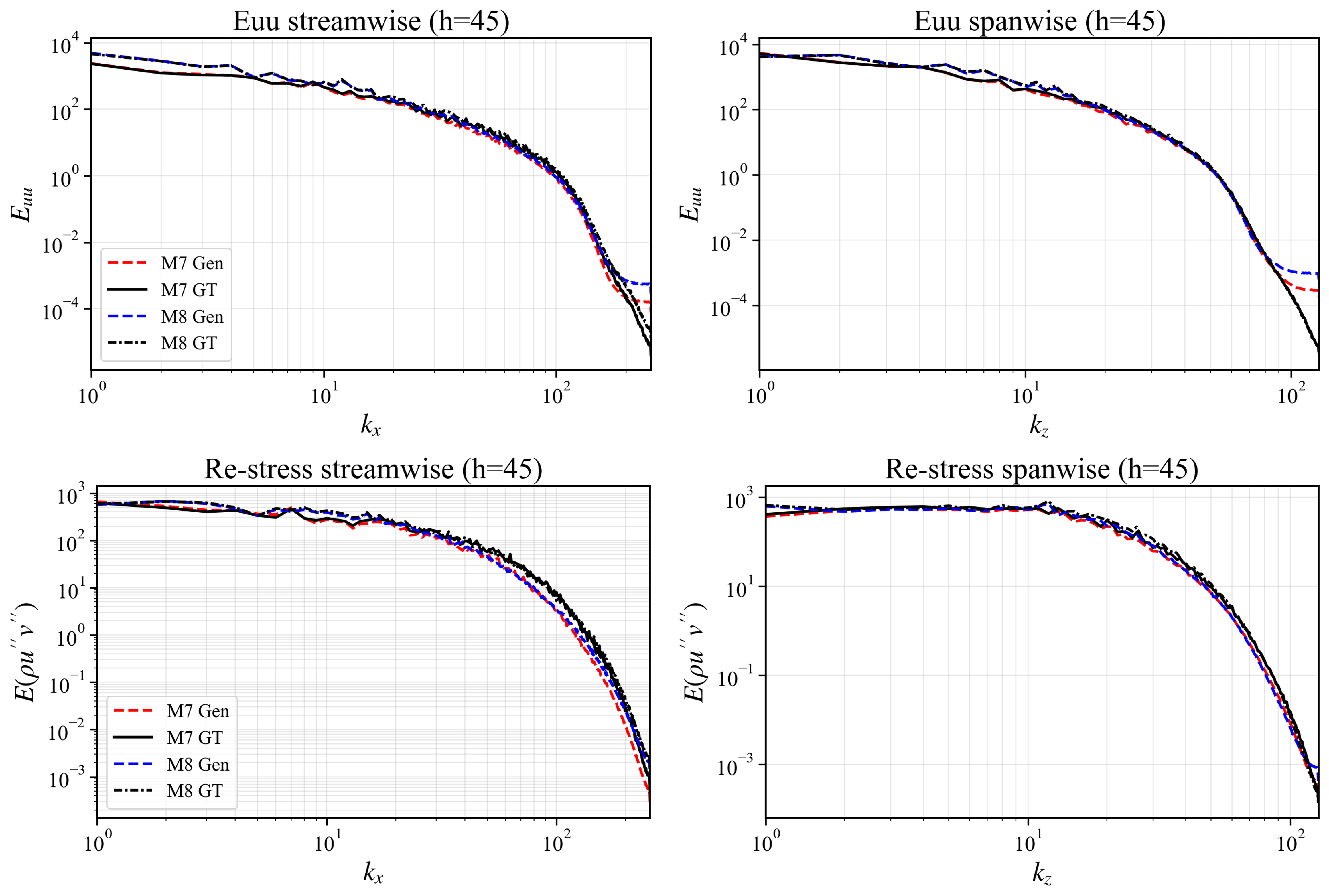}
\caption{One-dimensional energy spectra of $u''$ (top row) and Reynolds shear stress (bottom row) in the streamwise (left) and spanwise (right) directions at $y/\delta = 0.1128$ for $M=7$ and $M=8$: generated versus DNS. The EDM+BSP model matches the DNS spectra across the full wavenumber range at both Mach numbers, indicating that the spectral fidelity afforded by the bounded BSP loss extends across the training Mach range.}
\label{fig:spectra_M78}
\end{figure}

Figure~\ref{fig:spectra_M78} shows the corresponding one-dimensional energy and Reynolds-stress spectra at $y/\delta=0.1128$ for $M=7$ and $M=8$. As at $M=6$, the EDM+BSP model matches the DNS spectra across the full wavenumber range at both Mach numbers, demonstrating that the spectral fidelity afforded by the bounded BSP loss extends to the other training conditions.

% \begin{figure}
% \centering
% \includegraphics[width=\textwidth]{Figure/contour_M7_near_wall.png}
% \caption{Instantaneous wall-parallel contours of all six flow variables at two near-wall heights for $M=7$: generated (left) versus DNS (right).}
% \label{fig:contour_M7_near_wall}
% \end{figure}

\begin{figure}
\centering
\includegraphics[width=\textwidth]{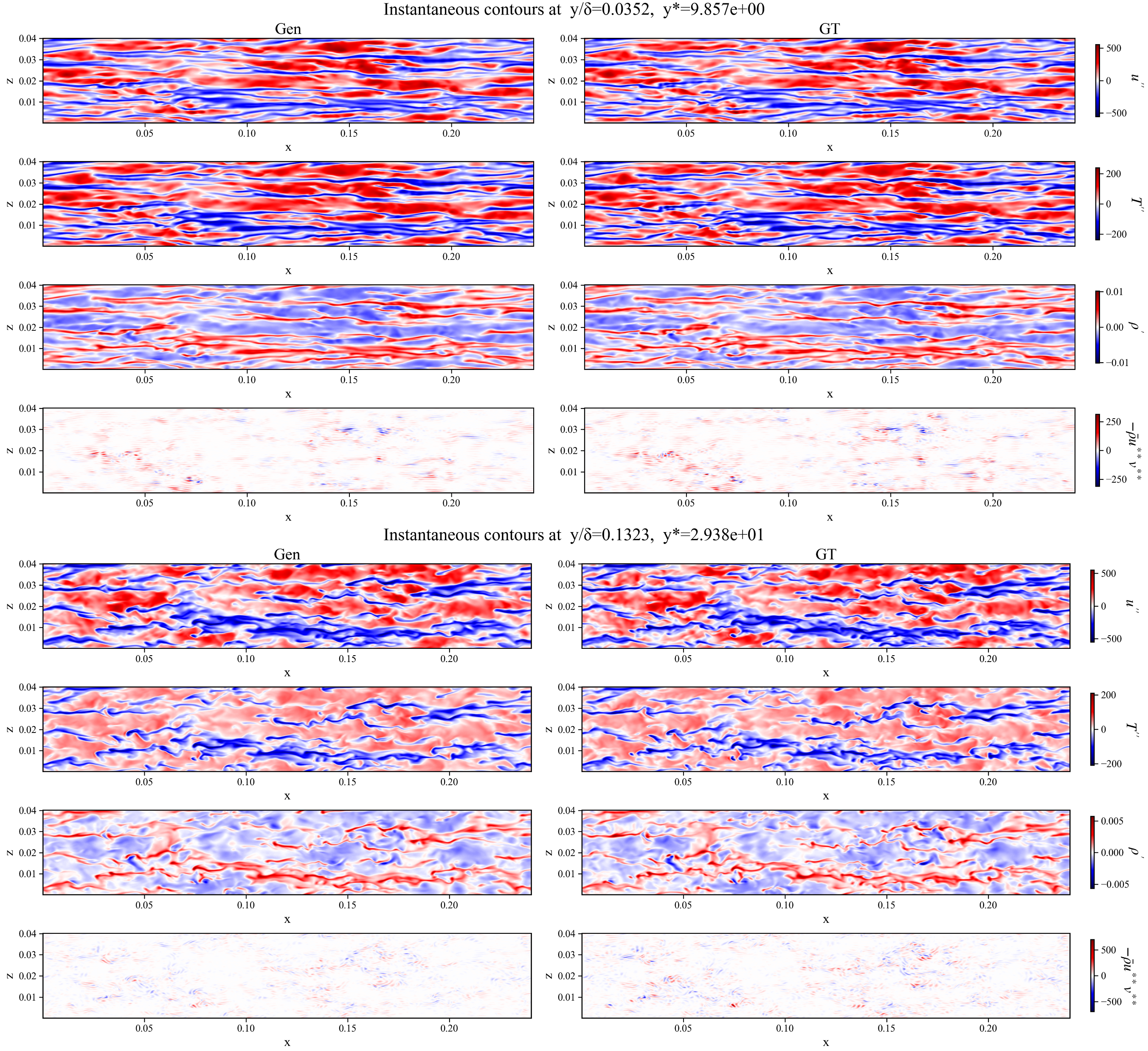}
\caption{Instantaneous wall-parallel contours at two near-wall heights for $M=8$: generated (left) versus DNS (right). Rows from top to bottom: Favre velocity fluctuation $u''$, Favre temperature fluctuation $T''$, Reynolds density fluctuation $\rho'$, and the second-order residual-velocity contribution to the SFS shear stress $-\bar{\rho}\,u^{**}v^{**}$ obtained from a sharp spectral filter with cut-off wavenumbers $(k_x, k_z) = (32, 16)$. The model reproduces the elongated streamwise streaks, the inter-variable correlations, and the small-scale SFS patterns with fidelity comparable to the $M=6$ case (Figs.~\ref{fig:contour_M6_near_wall_volume}- \ref{fig:contour_M6_near_wall_2_volume}).}
\label{fig:contour_M8_near_wall}
\end{figure}

Figure ~\ref{fig:contour_M8_near_wall} shows instantaneous wall-parallel contours at two near-wall heights for $M=8$. The model reproduces the elongated streamwise streaks and inter-variable correlations with fidelity comparable to the $M=6$ case (Figs.~\ref{fig:contour_M6_near_wall_volume}-\ref{fig:contour_M6_near_wall_2_volume}), confirming that the Mach-conditioning mechanism enables a single model to capture Mach-dependent flow structure.

\begin{figure}
\centering
\includegraphics[width=\textwidth]{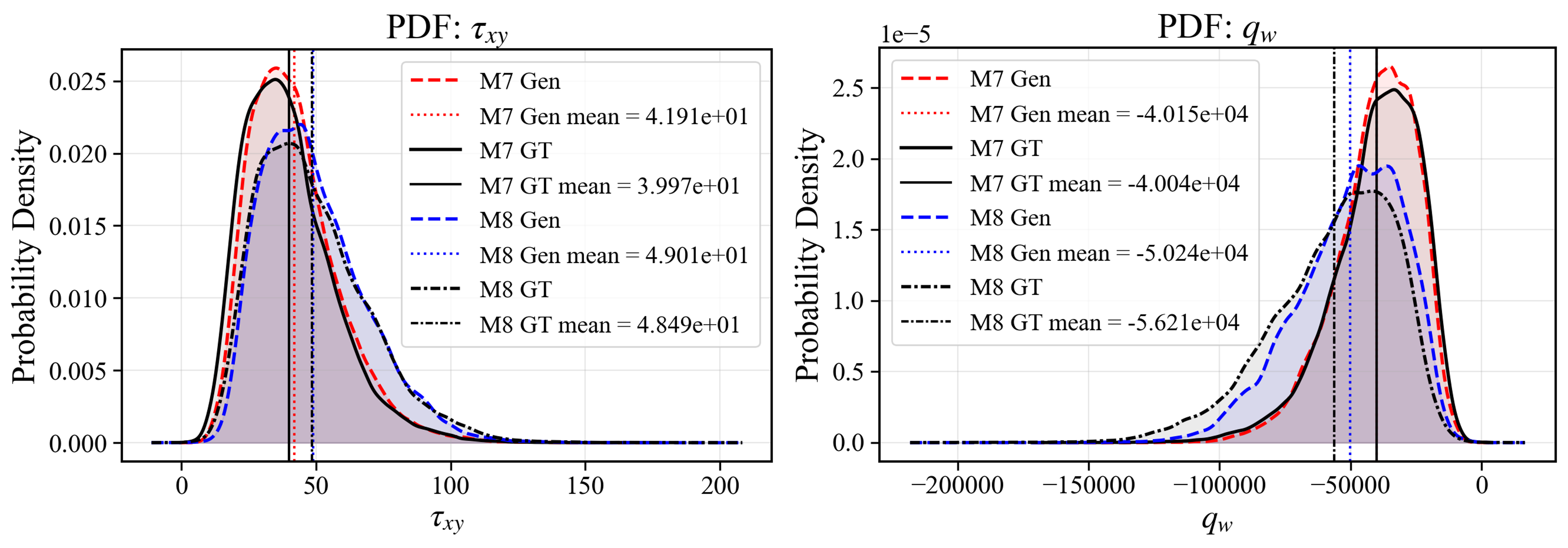}
\caption{PDFs of streamwise shear stress $\tau_{xy}$ (left) and wall heat flux $q_w$ (right) at $M=7$ and $M=8$: generated versus DNS. Mean values for each distribution are reported in the legend. The generated PDFs match the DNS distributions at both Mach numbers, with mean values in quantitative agreement.}
\label{fig:tau_xy_q_bot_M78}
\end{figure}

Figure~\ref{fig:tau_xy_q_bot_M78} compares PDFs of streamwise wall shear stress $\tau_{xy}$ and wall heat flux $q_w$ at $M=7$ and $M=8$. The generated PDFs match the DNS distributions closely at both Mach numbers, with quantitative agreement in the mean values reported in the legends.

\begin{figure}
\centering
\includegraphics[width=\textwidth]{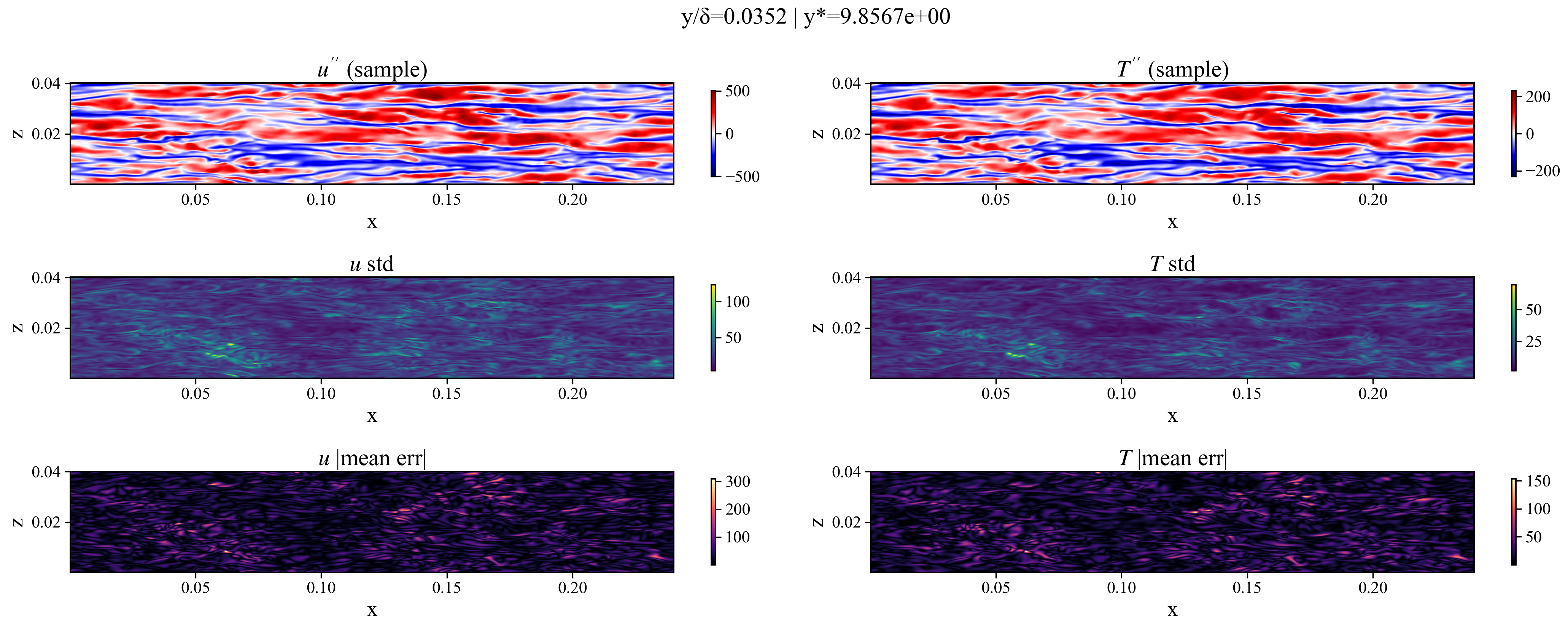}
\caption{Spatial uncertainty analysis at $y/\delta = 0.0352$ ($y^* \approx 9.9$), $M=8$. Top: instantaneous fluctuation fields $u''$ (left) and $T''$ (right) from a single generated sample. Middle: ensemble predictive standard deviation. Bottom: absolute mean error for the streamwise velocity and temperature. The predictive standard deviation again concentrates along inter-streak boundaries and spatially correlates with the absolute mean error, reproducing the behavior observed at $M=6$.}
\label{fig:UQcontour_M8_near_wall}
\end{figure}

\begin{figure}
\centering
\includegraphics[width=\textwidth]{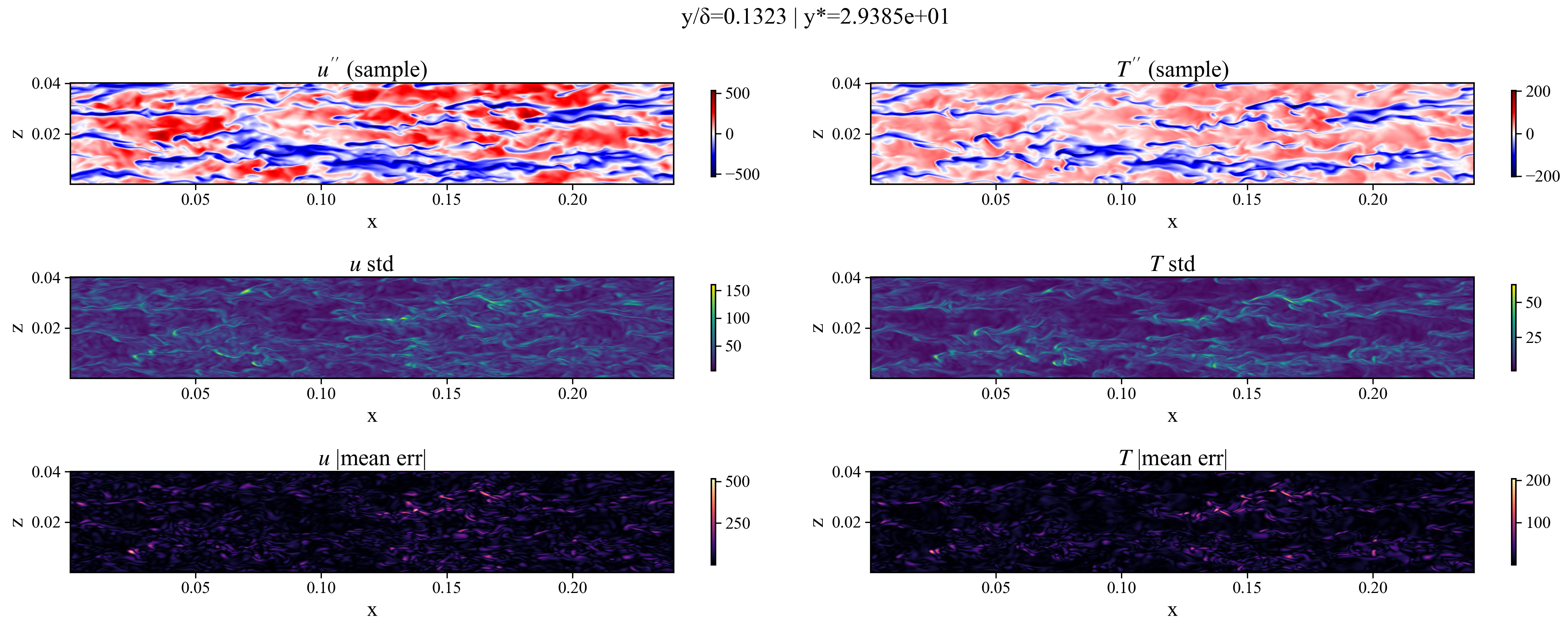}
\caption{Spatial uncertainty analysis at $y/\delta = 0.1128$ ($y^* \approx 29.4$), $M=8$. Top: instantaneous fluctuation fields $u''$ (left) and $T''$ (right) from a single generated sample. Middle: ensemble predictive standard deviation. Bottom: absolute mean error for the streamwise velocity and temperature.}
\label{fig:UQcontour_M8_near_wall_2}
\end{figure}

To verify that the physically meaningful uncertainty observed at $M=6$ extends across Mach numbers, Figs.~\ref{fig:UQcontour_M8_near_wall}-\ref{fig:UQcontour_M8_near_wall_2} show spatial uncertainty maps at two near-wall locations for $M=8$, in the same format as Figs.~\ref{fig:UQcontour_M6_near_wall}-\ref{fig:UQcontour_M6_near_wall_2}. The predictive standard deviation exhibits the same coherent spatial structure as at $M=6$: elevated uncertainty concentrates along inter-streak boundaries and spatially correlates with the absolute mean error, while the overall uncertainty decreases at the higher wall-normal location. This consistent behavior across the three training Mach numbers indicates that the physically meaningful uncertainty of the framework is a general property of the conditioning mechanism rather than specific to any single Mach number.

% \begin{figure}
% \centering
% \includegraphics[width=\textwidth]{Figure/UQcontour_M7_near_wall.png}
% \caption{Spatial uncertainty analysis at $y/\delta = 0.0352$ ($y^* \approx 10.6$), $M=7$. Top: instantaneous fluctuation fields $u''$ (left) and $T''$ (right) from a single generated sample. Middle: ensemble predictive standard deviation. Bottom: absolute mean error $|\bar{y}_\mathrm{gen} - y_\mathrm{DNS}|$.}
% \label{fig:UQcontour_M7_near_wall}
% \end{figure}
 
% \begin{figure}
% \centering
% \includegraphics[width=\textwidth]{Figure/UQcontour_M7_near_wall_2.png}
% \caption{Spatial uncertainty analysis at $y/\delta = 0.1128$ ($y^* \approx 32.6$), $M=7$. Top: instantaneous fluctuation fields $u''$ (left) and $T''$ (right) from a single generated sample. Middle: ensemble predictive standard deviation. Bottom: absolute mean error $|\bar{y}_\mathrm{gen} - y_\mathrm{DNS}|$.}
% \label{fig:UQcontour_M7_near_wall_2}
% \end{figure}

Although all three Mach numbers are seen during training, the training loss does not impose any compressibility-scaling constraint: the model is free to fit each Mach case independently via its $\phi_M$ embedding. We therefore use the Trettel-Larsson (TL) transformation \citep{trettel2016mean} as an independent diagnostic of whether the learned Mach conditioning encodes compressibility scaling consistently with known semi-empirical behavior. The TL transformation compensates for wall-normal variations in density and viscosity by defining a transformed velocity
\begin{equation}
    u_{TL}^+ = \int_0^{u^+} \left(\frac{\langle\rho\rangle}{\langle\rho_{bot}\rangle}\right)^{1/2}
    \left[1 + \frac{1}{2}\frac{1}{\langle\rho\rangle}\frac{\partial\langle\rho\rangle}{\partial y}\,y
    - \frac{1}{\langle\mu\rangle}\frac{\partial\langle\mu\rangle}{\partial y}\,y\right] \mathrm{d}u^+,
\end{equation}
plotted against the semi-local wall unit $y^* = \langle\rho\rangle u_\tau^* y / \langle\mu\rangle$, where $u_\tau^* = \sqrt{\tau_w / \langle\rho\rangle}$~\citep{huang1995compressible}. This transformation has been shown to collapse mean velocity profiles effectively in compressible Couette flows up to Mach 5~\citep{yao2023study}, and~\citet{toki2024sub} demonstrated that the collapse extends to hypersonic Couette flow at $M=6$ to $8$ using the same DNS dataset employed here.

\begin{figure}
\centering
\includegraphics[width=0.75\textwidth]{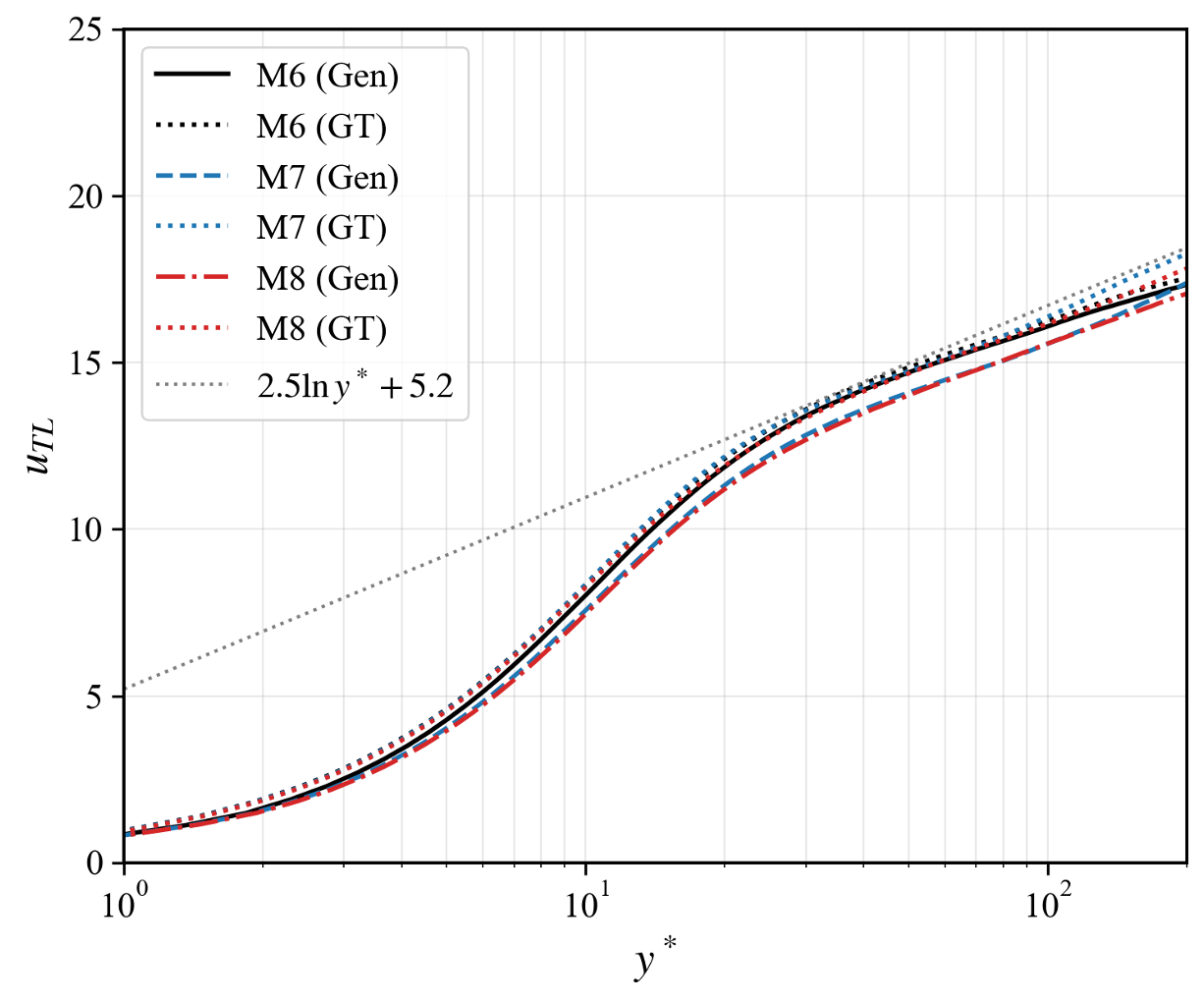}
\caption{Trettel-Larsson-transformed mean velocity profiles at $M=6,7,8$ with the reference log law $2.5\ln y^* + 5.2$. The transformed profiles from the generative model collapse across the three Mach numbers onto the log law, despite the absence of any compressibility-scaling constraint in the training loss.}
\label{fig:uTL_collapse}
\end{figure}

Figure~\ref{fig:uTL_collapse} shows that the TL-transformed mean streamwise velocity profiles from the generative model collapse across $M=6,7,8$ and agree closely with the standard log law $u_{TL}^+ = 2.5\ln y^* + 5.2$. Since \citet{toki2024sub} already established that the DNS profiles collapse under this transformation, the key observation here is that the generated profiles exhibit the same collapse despite the absence of any compressibility-scaling constraint in the training loss. This suggests that the learned conditioning is compatible with compressibility scaling of wall turbulence statistics.

\subsection{Generalization to unseen flow conditions}
\label{subsec:held-out}
The results above establish the in-distribution performance of the unified Mach-conditioned model at the three training Mach numbers. We now probe two complementary aspects of generalization of our framework. The first is interpolation within the trained Mach range: the model is retrained on $M=6$ and $M=8$ only, with $M=7$ withheld and used as a test case. The second is extrapolation in Reynolds number: the jointly-trained $M=6,7,8$ model is evaluated on an independent $M=8$ DNS at a substantially higher friction Reynolds number than any case seen during training (Toki et al.\ 2024, hereafter referred to as M8-R1), without any Reynolds-number conditioning. These two tests are designed to isolate generalization along the two principal axes spanned by the dataset, and the contrast between them clarifies which aspects of the underlying turbulent structure are captured by the present conditioning mechanism and which remain outside its current scope.

\subsubsection{Mach number interpolation: unseen $M = 7$}
For this test, the EDM+BSP model is retrained from scratch on $M=6$ and $M=8$ only, with all other architectural and training settings unchanged. Full-volume samples at the withheld Mach number are generated with the same soft overlap inpainting composition strategy introduced in Sec.~\ref{subsec:fullvolume}, with the Mach embedding $\phi_M$ evaluated at $M=7$ at inference time, so that the model must interpolate between the two training Mach numbers through the learned conditioning rather than reproduce a directly trained case.

\begin{figure}
\centering
\includegraphics[width=\textwidth]{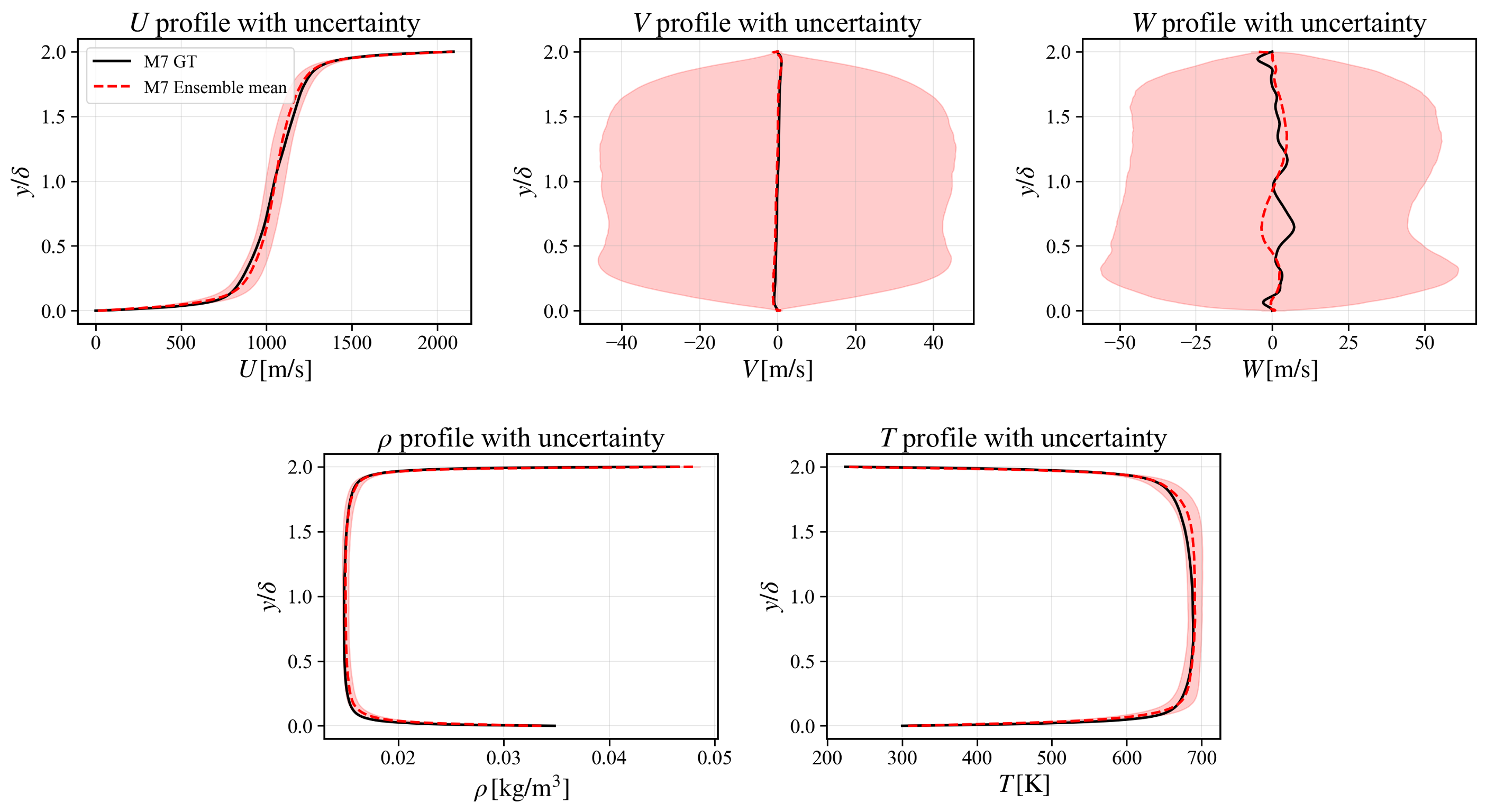}
\caption{Ensemble mean profiles (dashed) and $\pm 2\sigma$ uncertainty bands (shaded) of $U$, $V$, $W$, $\rho$, and $T$ at the unseen Mach number $M=7$: generated (EDM+BSP retrained on $M=6,8$ only, with soft overlap inpainting) versus DNS (solid). The ensemble mean tracks the DNS profile across the entire channel for all five variables, with tight uncertainty bands for $U$, $\rho$, and $T$ that enclose the ground truth.}
\label{fig:mean_profiles_M7_uq_heldout}
\end{figure}

Figure~\ref{fig:mean_profiles_M7_uq_heldout} shows ensemble mean profiles and $\pm 2\sigma$ uncertainty bands of $U$, $V$, $W$, $\rho$, and $T$ at the unseen Mach number. The ensemble mean closely tracks the DNS profile across the entire channel, with tight bands for $U$, $\rho$, and $T$ that enclose the ground truth, and broader bands for the fluctuation-dominated $V$ and $W$ components, in the same pattern observed at the training Mach numbers (Figs.~\ref{fig:mean_profiles_M6_uq} and \ref{fig:mean_profiles_M78_uq}). The first-order structure of the boundary layer is therefore recovered at the unseen Mach number to a level visually comparable to the in-distribution cases.

\begin{figure}
\centering
\includegraphics[width=\textwidth]{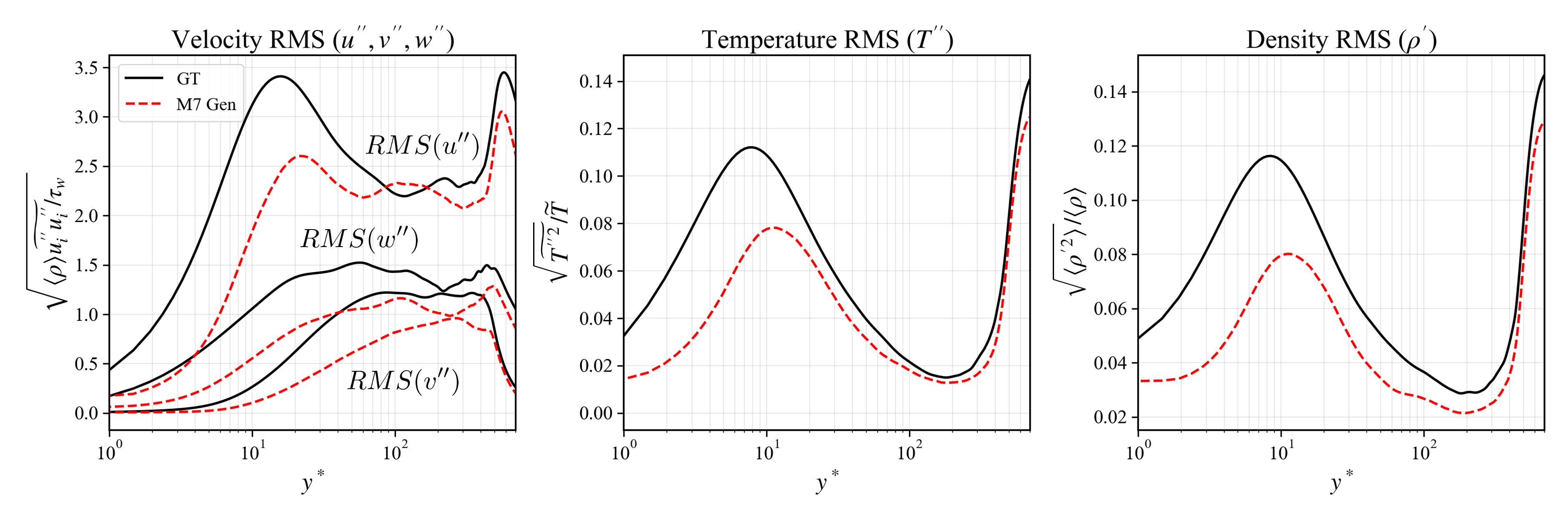}
\caption{RMS profiles of velocity components $u''$, $v''$, $w''$ (left), temperature (middle), and density (right) at the unseen Mach number $M=7$: generated versus DNS (GT).  The wall-normal shape of all profiles is preserved, but peak amplitudes are systematically underestimated below $y^* \approx 20$, most prominently for $u''$.}
\label{fig:RMS_profiles_M7_heldout}
\end{figure}

Figure~\ref{fig:RMS_profiles_M7_heldout} compares RMS profiles of the three velocity components, temperature, and density against DNS. The qualitative shape of all profiles is preserved: the generated curves exhibit the expected near-wall peak followed by a monotonic outer-layer decay, and the relative ordering of $u''$, $v''$, and $w''$ matches the DNS. The peak amplitudes, however, are systematically underestimated. The streamwise velocity fluctuation $u''$ shows the largest discrepancy, with the near-wall peak ($y^* \approx 10$--$20$) reaching only about 75\% of the DNS value; the wall-normal and spanwise components $v''$ and $w''$, as well as the temperature and density RMS, exhibit a milder but consistent reduction in peak amplitude and a small shift of the peak location toward the wall. The deviation is most pronounced in the inner layer ($y^* \lesssim 20$) and attenuates with wall distance, indicating that the withheld Mach setting preserves the wall-normal organization of fluctuations but loses some of their near-wall energy content.

\begin{figure}
\centering
\includegraphics[width=\textwidth]{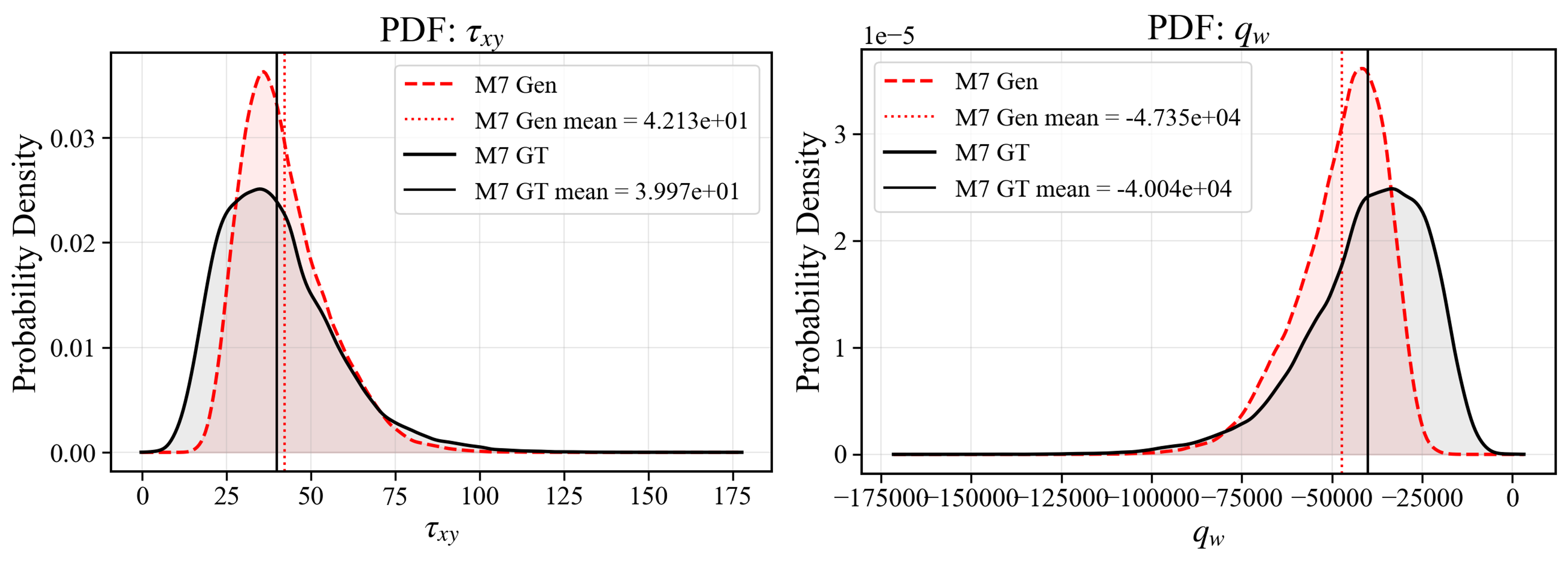}
\caption{PDFs of  streamwise shear stress $\tau_{xy}$ (left) and wall heat flux $q_w$ (right) at the withheld Mach number $M=7$: generated versus DNS. Mean values for each distribution are reported in the legend.  The generated distributions are noticeably sharper than the DNS, with overestimated peak density and underestimated high-magnitude tails.}
\label{fig:tau_xy_q_bot_M7_heldout}
\end{figure}

Figure~\ref{fig:tau_xy_q_bot_M7_heldout} compares PDFs of wall heat flux $q_w$ and streamwise shear stress $\tau_{xy}$ between the generated samples and the DNS. The generated PDFs are approximately centered on the DNS distributions: the peak locations coincide and the ensemble mean values reported in the legend are in reasonable quantitative agreement with the DNS. The shape of the generated PDFs, however, is noticeably sharper than the DNS in both cases: the peak density is overestimated by roughly $40$\% and the high-magnitude tails are correspondingly underestimated. This over-concentration is consistent with the underestimated near-wall fluctuation amplitudes observed in Fig.~\ref{fig:RMS_profiles_M7_heldout}, since wall fluxes are dictated by near-wall velocity and temperature gradients whose tail behavior is set precisely by the high-amplitude excursions that are not fully captured at the withheld Mach number.

Taken together, these results indicate that the learned Mach conditioning is not a discrete look-up across the training Mach numbers but a continuous representation that supports interpolation of first-order statistics: mean profiles and the bulk wall-normal organization of fluctuations are recovered at the unseen Mach number to a level comparable with the in-distribution cases, and the wall-quantity means are reproduced quantitatively. Higher-order near-wall statistics, by contrast, are systematically attenuated: peak RMS amplitudes are underestimated below $y^* \approx 20$, and the resulting wall-quantity PDFs are over-peaked with underestimated tails. Two factors plausibly contribute to this attenuation, both of which are tied to the training setup of this particular test. First, the absence of $M=7$ from the training set removes direct supervision at the test Mach number, so the buffer- and viscous-sublayer structure must be reconstructed entirely from cross-Mach interpolation. Second, and arguably more limiting, the training set in this test contains only two Mach numbers, which is an inherently sparse sampling of the Mach axis: two anchor points convey no information about how near-wall fluctuation amplitudes vary nonlinearly between them, and intermediate behavior is therefore reconstructed solely from whatever interpolation bias is built into the embedding. A denser sampling along the Mach axis would be expected to relax this constraint substantially, and is a natural follow-up to the present experiment. We also note that the in-distribution evaluation in Secs.~\ref{subsec:training_case} and \ref{subsec:multimach}, where each of $M=6$, $7$, and $8$ is directly represented in the training data, recovers these same second-order near-wall statistics with high fidelity, which is consistent with this interpretation.

\subsubsection{Reynolds number extrapolation: high-$Re_\tau$ case} We next assess the behavior of the jointly-trained $M=6,7,8$ model when applied to flow conditions that lie outside the Reynolds-number range represented during training. The test case (M8-R1, Toki et al.\ 2024) is an independent DNS of hypersonic Couette flow at $M=8$ with $Re_\tau = 1278$, approximately $1.8$ times the friction Reynolds number of the corresponding $M=8$ training case ($Re_\tau = 713$) and well above the largest training value across all three training Mach numbers ($Re_\tau \in [621, 713]$). Full-volume samples at this condition are generated with the soft overlap inpainting composition described in Sec.~\ref{subsec:fullvolume}. Reynolds number is not an explicit conditioning input in the present model, so this test probes how the framework responds when only the Mach embedding is matched while the underlying turbulent state corresponds to a regime well outside the training envelope.

\begin{figure}
\centering
\includegraphics[width=\textwidth]{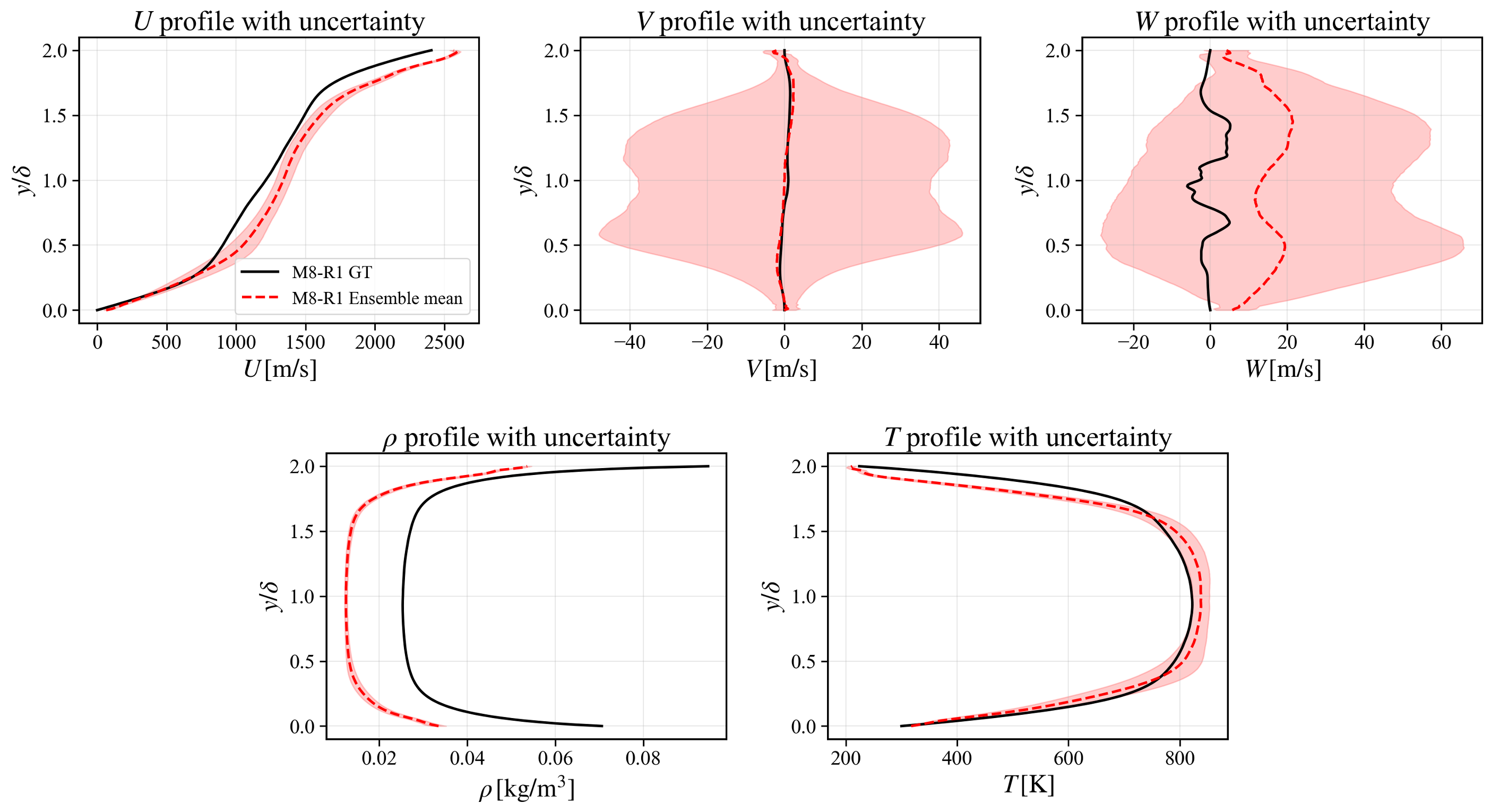}
\caption{Ensemble mean profiles (dashed) and $\pm 2\sigma$ uncertainty bands (shaded) of $U$, $V$, $W$, $\rho$, and $T$ at the M8-R1 test condition ($M=8$, $Re_\tau = 1278$, approximately $1.8\times$ the largest training $Re_\tau$): generated (EDM+BSP trained on $M=6,7,8$ with soft overlap inpainting) versus DNS (solid). The mean profiles exhibit a systematic offset from the DNS, and the predictive uncertainty bands do not inflate sufficiently to encompass the ground truth across most of the channel.}
\label{fig:mean_profiles_M8_R1_uq_heldout}
\end{figure}

\begin{figure}
\centering
\includegraphics[width=\textwidth]{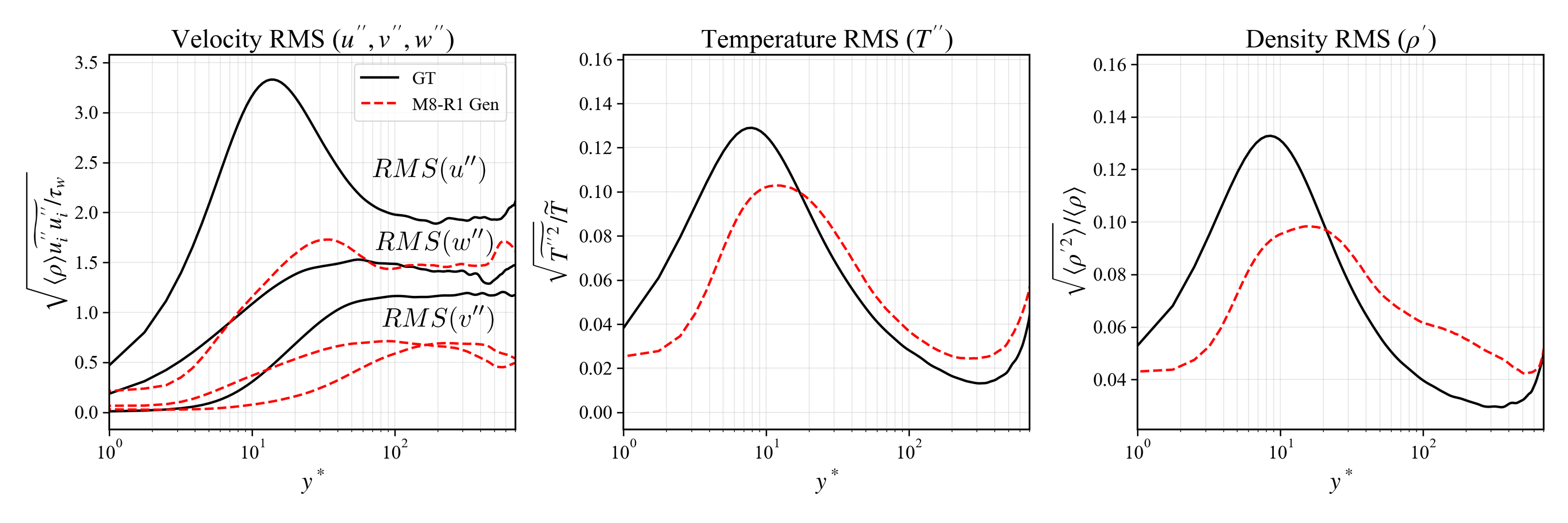}
\caption{RMS profiles of velocity components $u''$, $v''$, $w''$ (left), temperature (middle), and density (right) at the M8-R1 test condition: generated versus DNS (GT). The qualitative profile shape is retained, but peak locations and amplitudes do not align with the DNS, reflecting the absence of Reynolds-number conditioning in the trained model.}
\label{fig:RMS_profiles_M8_R1_heldout}
\end{figure}

Figure~\ref{fig:mean_profiles_M8_R1_uq_heldout} shows ensemble mean profiles and $\pm 2\sigma$ uncertainty bands of $U$, $V$, $W$, $\rho$, and $T$ at the M8-R1 condition. In contrast to the withheld Mach test, the generated mean profiles exhibit a systematic offset from the DNS: the streamwise velocity boundary layer is recovered with a shifted profile shape, and analogous discrepancies are visible in the density and temperature fields. The predictive uncertainty bands do not inflate sufficiently to encompass the ground truth in most of the channel, indicating that the model is confidently biased toward the Reynolds-number regime of the training set rather than spontaneously broadening its posterior to reflect the regime mismatch. Figure~\ref{fig:RMS_profiles_M8_R1_heldout} supports this picture for the second-order statistics: the generated RMS profiles of velocity, temperature, and density retain the qualitative shape of the training-Reynolds-number profiles, but the peak positions and amplitudes do not align with the M8-R1 DNS, particularly in the buffer and log layers where the wall-normal scaling is most strongly Reynolds-number dependent.

\begin{figure}
\centering
\includegraphics[width=\textwidth]{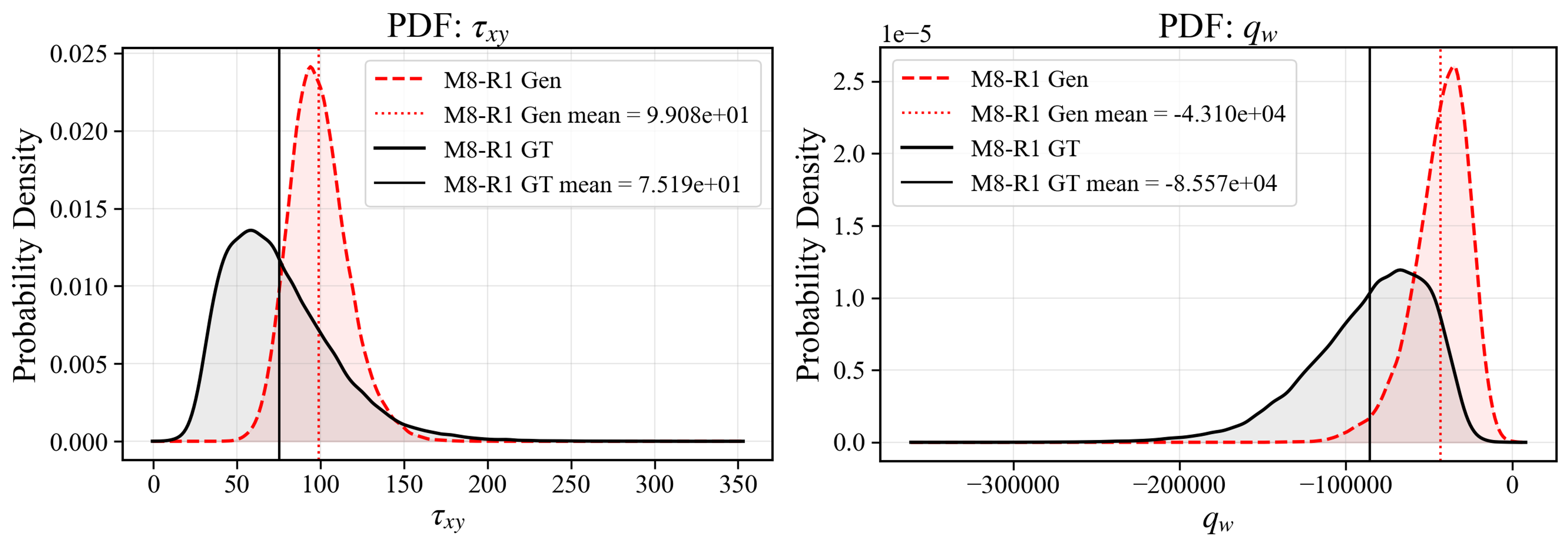}
\caption{PDFs of streamwise shear stress $\tau_{xy}$ (left) and wall heat flux $q_w$ (right) at the M8-R1 test condition: generated versus DNS. Mean values for each distribution are reported in the legend. Both PDFs are systematically shifted relative to the DNS, consistent with the near-wall gradient mismatch inferred from the profile-level diagnostics.}
\label{fig:tau_xy_q_bot_M8_R1_heldout}
\end{figure}

The corresponding wall-quantity PDFs in Fig.~\ref{fig:tau_xy_q_bot_M8_R1_heldout} are shifted relative to the DNS for both $q_w$ and $\tau_{xy}$, consistent with the near-wall gradient mismatch inferred from the profile-level diagnostics.

These observations are consistent with the design of the present framework rather than evidence of a failure of the conditional-diffusion formulation itself: the model is conditioned on Mach number but not on Reynolds number, and the M8-R1 friction Reynolds number lies well outside the range spanned by the training data. The framework therefore generalizes along the explicitly-conditioned compressibility axis but not along the unconditioned Reynolds-number axis, which delineates the operating envelope of the model in its current form. This represents the limitations of our data-driven approach, reflecting the need for optimal experimental design for parameteric generalization.

Read together, the two generalization experiments in this section point to a common operating principle: the fidelity of the reconstruction at unseen flow conditions is governed primarily by the density of training coverage along the relevant physical axis. In Sec.~4.3.1, sparse two-point sampling of the Mach axis was sufficient to interpolate mean structure but attenuated near-wall fluctuation amplitudes at the intermediate Mach number; here, the complete absence of Reynolds-number coverage, both in the conditioning and in the training data, produces a substantially more severe degradation. Both observations are consistent with the broader expectation that conditional generative models extrapolate poorly beyond their training support and interpolate accurately only when the parametric grid is sufficiently dense. They jointly motivate a single concrete extension, which we leave to future work: training the same subdomain-wise diffusion framework on a denser two-dimensional grid of Mach and Reynolds numbers, with the conditioning extended to include $Re_\tau$ (or an equivalent outer-scale parameter) alongside the existing Mach embedding. Such an extension would directly address the two limitations identified above and is expected to substantially broaden the Mach Reynolds envelope over which the framework can be deployed.

%==========================================================
\section{Conclusion}
\label{sec:conclusions}
 
We have introduced a multiscale probabilistic reconstruction framework for hypersonic Couette flow built around two complementary ingredients: a subdomain-wise conditional diffusion generation strategy that samples the full volume as a collection of height- and Mach-conditioned wall-normal subdomains stitched together by soft overlap inpainting, and a spectral-binning training loss that explicitly supervises binned one-dimensional energy spectra in a bounded, numerically stable form compatible with the EDM noise schedule. The framework adapts and extends ideas from recent patch-trained and joint-diffusion work in the image and medical-imaging communities to the specific demands of compressible wall-bounded turbulence: physically heterogeneous wall-normal subdomains, a variance-preserving inpainting composition (in place of the arithmetic or Gaussian averaging common in prior joint-diffusion frameworks), and explicit multiscale spectral supervision. Together, these components make subdomain-wise generation tractable, keep inter-subdomain continuity from collapsing small scale variance, and sharpen high-wavenumber spectral content of the generated fields.
 
Across DNS benchmarks of hypersonic Couette flow at $M=6$, $7$, and $8$, the single jointly-trained Mach-conditioned framework recovers dominant streaky coherent structures, matches one-dimensional energy spectra and Reynolds stress spectra, reproduces PDFs of wall shear stress and heat flux, and supplies physically plausible uncertainty estimates on mean profiles, Reynolds-stress profiles, two-point correlations, and wall quantities, demonstrating that a single unified model can represent the three training Mach numbers through its conditioning mechanism. Moreover, the Trettel-Larsson-transformed mean velocity profiles collapse across the three Mach numbers onto the standard log law, indicating that the learned Mach conditioning is consistent with the known compressibility scaling of wall-bounded turbulence.

These results also point to several directions for future work. Building on the generalization experiments with $M=7$ and high-$Re_\tau$ in Sec.~\ref{subsec:held-out}, a natural next step is to train the same framework on a denser two-dimensional grid of Mach and Reynolds numbers, with the conditioning extended to include $Re_\tau$ (or an equivalent outer-scale parameter). Such an extension would directly address the two generalization-axis limitations identified in our assessments and would also allow a more systematic study of how the Trettel-Larsson collapse observed for the generated profiles transfers to unseen Mach-Reynolds regimes. In parallel, the reconstruction model should be coupled with CFD solvers, for example as a dynamic near-wall boundary condition for wall-modeled LES of hypersonic flows, where its \emph{a posteriori} effects on solver stability, accuracy, and computational cost can be assessed. A broader extension is to move beyond the idealized Couette configuration and apply the framework to turbulent boundary layers over curved and three-dimensional geometries, including non-equilibrium effects, shock/boundary-layer interactions, and transitional regimes. Finally, data-efficient conditioning remains an important practical question: future studies should examine how reliably the posterior can be sampled when the outer-layer information is replaced by sparser observations, such as wall-only sensors or low-fidelity LES interface data.

\backsection[Acknowledgements]{This research used resources of the Argonne Leadership Computing Facility, which is a U.S. Department of Energy (DOE) Office of Science User Facility operated under Contract DE-AC02-06CH11357.}

\backsection[Funding]{We acknowledge support from ARO cooperative agreement W911NF-25-2-0183 and an ARO ECP award from the Program `Modeling of Complex Systems' (PM - Dr. Rob Martin).}

\backsection[Declaration of interests]{The authors report no conflict of interest.}

% \backsection[Data availability statement]{The data that support the findings of this study are openly available in [repository name] at http://doi.org/[doi], reference number [reference number]. See JFM's \href{https://www.cambridge.org/core/journals/journal-of-fluid-mechanics/information/journal-policies/research-transparency}{research transparency policy} for more information}

% \backsection[Author ORCIDs]{Authors may include the ORCID identifers as follows.  F. Smith, https://orcid.org/0000-0001-2345-6789; B. Jones, https://orcid.org/0000-0009-8765-4321}

\appendix

\section{EDM Training and Sampling Details}
\label{app:edm}

This appendix collects implementation details of the conditional EDM framework introduced in Sec.~\ref{subsec:diffusion}, including the noise schedule, the network architecture, the two-phase training procedure, and the deterministic sampler used at inference.

\subsection{Noise schedule and preconditioning}

The training noise level $\sigma$ is sampled from a log-normal distribution,
\begin{equation}
    \ln\sigma \sim \mathcal{N}(P_\text{mean}, P_\text{std}^2),
\end{equation}
following~\citet{karras2022elucidating}. At inference, the deterministic sampler integrates the probability-flow ODE of Eq.~\eqref{eq:prob_flow_ode} from $\sigma_\text{max}$ down to $\sigma_\text{min}$, with the discretized noise levels following the EDM Karras schedule
\begin{equation}
    \sigma_i = \left(\sigma_\text{max}^{1/\rho} + \frac{i}{N_d - 1}\bigl(\sigma_\text{min}^{1/\rho} - \sigma_\text{max}^{1/\rho}\bigr)\right)^{\rho},
    \qquad i = 0, 1, \ldots, N_d - 1,
\end{equation}
with $\rho = 7$. Table~\ref{tab:edm_hyperparams} summarizes the key hyperparameters.

\begin{table}
\centering
\begin{tabular}{lc}
\hline
Parameter & Value \\
\hline
$\sigma_\text{min}$ & $0.002$ \\
$\sigma_\text{max}$ & $80.0$ \\
$\sigma_\text{data}$ & $1.0$ \\
$\rho$ (sampler schedule) & $7$ \\
$P_\text{mean}$ (Phase 2) & $-1.0$ \\
$P_\text{std}$ (Phase 2) & $1.2$ \\
$N_d$ (sampler steps) & $18$ \\
$\beta$ (BSP weight) & $10^{-3}$ \\
$N_k$ (BSP wavenumber bins per direction) & $32$ \\
\hline
\end{tabular}
\caption{EDM and BSP hyperparameters used in this work.}
\label{tab:edm_hyperparams}
\end{table}

\subsection{Denoiser architecture}

The denoiser $F_\theta$ is implemented as a SongUNet~\citep{song2020score} with periodic (circular) padding to respect the streamwise and spanwise homogeneity of the flow. The base channel width is $128$, with channel multipliers $(1, 2, 2, 2, 2)$ across five resolution levels and four residual blocks per resolution. Multi-head self-attention is applied at the resolution corresponding to the smallest feature map size ($32 \times 16$). Group normalization is used throughout, and a dropout rate of $0.1$ is applied within residual blocks. The total number of trainable parameters is approximately $80$ million.

The conditioning embedding
\begin{equation}
    \mathbf{e} = \mathbf{e}_\sigma + \phi_h(h) + \phi_M(M)
\end{equation}
is constructed as follows. The noise-level embedding $\mathbf{e}_\sigma$ is a sinusoidal positional encoding of $c_\text{noise}(\sigma) = \tfrac{1}{4}\ln\sigma$ followed by a two-layer MLP. The height embedding $\phi_h$ takes the normalized wall-normal height $h \in [0, 1]$ and the Mach embedding $\phi_M$ takes the integer Mach number $M \in \{6, 7, 8\}$; both are passed through independent linear projections to the same dimensionality as $\mathbf{e}_\sigma$. The combined embedding $\mathbf{e}$ is broadcast to every residual block via adaptive group normalization (AdaGN), where it produces per-block scale and shift parameters that modulate the feature maps as a function of noise level, wall-normal position, and flow condition.

\subsection{Training data}

Each training example pairs a top-wall boundary slice $\mathbf{x}_b \in \mathbb{R}^{6 \times 512 \times 256}$ with a subdomain $\mathbf{y}_h \in \mathbb{R}^{6 \times N_s \times 512 \times 256}$ of $N_s = 4$ consecutive wall-normal slices starting at a wall-normal index sampled from an exponential decay distribution biased toward near-wall positions, drawn from the same DNS snapshot. The Mach number $M \in \{6, 7, 8\}$ is supplied as a scalar conditioner. Both the boundary slice and the subdomain are normalized to zero mean and unit variance using per-channel statistics computed over the full training set; sampled subdomains are de-normalized before composition into the full volume. At each Mach number, the dataset of DNS snapshots is split $80/20$ between training and held-out evaluation, with the $20\%$ evaluation snapshots withheld from training and used exclusively for the sampling and statistical evaluation reported in Sec.~\ref{sec:Results}.

\subsection{Two-phase training schedule}

Training is performed in two phases. \emph{Phase 1} optimizes the EDM denoising loss alone (Eq.~\ref{eq:edm_loss}); \emph{Phase 2} resumes from the EMA weights of Phase 1 and adds the bounded BSP loss, training with the composite objective of Eq.~\eqref{eq:composite_loss}. Both phases use AdamW with $(\beta_1, \beta_2) = (0.9, 0.95)$, $\epsilon = 10^{-6}$, learning rate $2 \times 10^{-4}$, and weight decay $10^{-5}$; the learning rate schedule consists of a linear warm-up over the first $1000$~kimg followed by cosine annealing down to $1\%$ of the maximum value after a total of 23000 kimg. An exponential moving average of the network weights is maintained with a half-life of $50$~kimg and used for all evaluation and sampling.

The two phases differ only in the loss function and in the noise distribution parameters, as summarized in Table~\ref{tab:phases}. The shift to a slightly more efficient noise distribution at the transition between Phase 1 and Phase 2 followed standard EDM practice for fine-tuning. Training is performed in distributed data-parallel fashion across $12$ GPUs on Aurora (ALCF, Intel XPU), with a local batch size of $4$ subdomains per GPU (global batch size of $48$).

\begin{table}
\centering
\begin{tabular}{lcc}
\hline
Setting & Phase 1 & Phase 2 \\
\hline
Loss & EDM only & EDM $+\,10^{-3} \cdot \mathcal{L}_\text{BSP}$ \\
$P_\text{mean}$ & $-0.5 \to -1.0$ & $-1.0$ \\
$P_\text{std}$ & $1.5 \to 1.2$ & $1.2$ \\
BSP wavenumber bins $N_k$ & --- & $32$ (linearly spaced) \\
Initialization & From scratch & From Phase 1 EMA weights \\
\hline
\end{tabular}
\caption{Training hyperparameters that differ between Phase 1 and Phase 2.}
\label{tab:phases}
\end{table}

\subsection{Sampling}

Inference uses Heun's second-order deterministic sampler over the discretized noise schedule, as recommended by~\citet{karras2022elucidating}. At each step, the denoiser is evaluated to predict the clean sample, followed by a corrector step that refines the update of the probability-flow ODE; this gives second-order accuracy. Sampling is performed without stochastic injection, so that the only stochasticity in the generated ensemble enters through the initial Gaussian noise $\mathbf{y}_{h,\sigma_\text{max}} \sim \mathcal{N}(\mathbf{0}, \sigma_\text{max}^2 \mathbf{I})$. For the uncertainty quantification results (Sec.~\ref{subsec:training_case} and Sec.~\ref{subsec:multimach}), an ensemble of multiple samples is generated for each conditioning $\mathbf{c}$ by drawing independent initial noise per sample.

\section{Soft Overlap Inpainting Algorithm}
\label{app:inpainting}

This appendix provides the pseudo-code for the soft overlap inpainting scheme described in Sec.~\ref{subsec:fullvolume}. The full wall-normal stack of $N_y = 192$ planes is generated as a sequence of overlapping subdomains of depth $N_s = 4$, traversed from the bottom wall toward the top wall with a sliding-window stride. The first subdomain advances by $\Delta h_0 = 2$ planes (\emph{first stride}); all subsequent subdomains advance by $\Delta h = 1$ plane (\emph{stride}), so that consecutive subdomains overlap on $N_s - \Delta h = 3$ planes. Under this configuration, the full volume of $192$ planes is assembled from $N_c = 189$ subdomains generated sequentially.

The first subdomain is sampled unconstrained from pure noise. Each subsequent subdomain is sampled with the soft overlap inpainting scheme: at every step of the reverse-diffusion process, the planes shared with the previously generated subdomain are softly blended toward the corresponding noised values from the previous subdomain. The blending coefficient $\alpha(\sigma_t)$ follows the cosine schedule of Eq.~\eqref{eq:alpha_schedule} with $\alpha_\text{start} = 0.5$ and $\alpha_\text{end} = 0$, monotonically decreasing along the reverse-diffusion direction. Algorithm~\ref{alg:inpainting} summarizes the procedure for generating the $(k{+}1)$-th subdomain given the previously generated $k$-th subdomain.

\begin{algorithm}
\caption{Soft overlap inpainting for the $(k{+}1)$-th subdomain}
\label{alg:inpainting}
\begin{algorithmic}[1]
\Require Conditioning $\mathbf{c} = (\mathbf{x}_b, h_{k+1}, M)$, denoiser $D_\theta$
\Require Previously generated subdomain $\hat{\mathbf{y}}_h^{(k)}$, overlap depth $N_o = N_s - \Delta h$
\Require Discretized noise schedule $\{\sigma_t\}_{t=0}^{N_d-1}$ with $\sigma_0 = \sigma_\text{max}$ and $\sigma_{N_d-1} = \sigma_\text{min}$
\State Extract overlap region $\mathbf{z}_\text{known}^{(k)} \gets \hat{\mathbf{y}}_h^{(k)}[\,:,\, \Delta h : N_s,\, :,\, : \,]$
\State Sample initial noise $\mathbf{y}_{h,\sigma_0} \sim \mathcal{N}(\mathbf{0}, \sigma_0^2 \mathbf{I})$
\For{$t = 1, 2, \ldots, N_d - 1$}
    \State $\mathbf{y}_{h,\sigma_t} \gets$ Heun ODE update from $\mathbf{y}_{h,\sigma_{t-1}}$ using $D_\theta(\,\cdot\,;\sigma_t, \mathbf{c})$
    \State Sample $\boldsymbol{\epsilon} \sim \mathcal{N}(\mathbf{0}, \sigma_t^2 \mathbf{I})$
    \State $\mathbf{z}_{\text{known},\sigma_t}^{(k)} \gets \mathbf{z}_\text{known}^{(k)} + \boldsymbol{\epsilon}$ \Comment{noise the known trajectory to level $\sigma_t$}
    \State $\mathbf{z}^{(k)}_{\sigma_t} \gets \mathbf{y}_{h,\sigma_t}[\,:,\, 0 : N_o,\, :,\, : \,]$ \Comment{model trajectory in overlap region}
    \State $\alpha_t \gets \tfrac{1}{4}\bigl(1 + \cos(\pi (t-1) / (N_d{-}2))\bigr)$
    \Comment{cosine schedule with $\alpha_\text{start} = 0.5$, $\alpha_\text{end} = 0$}
    \State $\mathbf{z}^{(k)}_{\sigma_t} \gets \alpha_t\, \mathbf{z}_{\text{known},\sigma_t}^{(k)} + (1 - \alpha_t)\, \mathbf{z}^{(k)}_{\sigma_t}$
    \State $\mathbf{y}_{h,\sigma_t}[\,:,\, 0 : N_o,\, :,\, : \,] \gets \mathbf{z}^{(k)}_{\sigma_t}$ \Comment{write blended overlap back}
\EndFor
\State \Return $\hat{\mathbf{y}}_h^{(k+1)} \gets \mathbf{y}_{h,\sigma_{N_d-1}}$
\end{algorithmic}
\end{algorithm}

After all $N_c$ subdomains are sampled, the full volume is assembled by concatenating the non-overlapping leading slices of each subdomain. With $\Delta h_0 = 2$ for the first subdomain and $\Delta h = 1$ thereafter, this procedure yields the full stack of $N_y = 192$ planes.

\bibliographystyle{jfm}
\bibliography{jfm}

\end{document}